\documentclass[A4]{article}

\usepackage{amsmath,amsfonts,amsthm,amssymb,amscd,color}
\usepackage{multirow}
\usepackage[dvipdfmx]{graphicx}
\usepackage[format=hang,font=footnotesize]{caption}

\theoremstyle{plain} \newtheorem{theorem}{Theorem}[section]

 \theoremstyle{definition}
 \theoremstyle{remark}
\newtheorem{remark}[theorem]{Remark} 
 
\newcommand{\R}{{\mathbb R}}

\newcommand{\Z}{{\mathbb Z}}

\newcommand{\N}{{\mathbb N}}

\newcommand{\C}{\mathbb{C}}

\def\({\left(}
\def\){\right)}
\def\<{\left\langle}
\def\>{\right\rangle}

\numberwithin{equation}{section}

\begin{document}

\title{Dynamics of solitons for nonlinear quantum walks}

\author {Masaya Maeda, Hironobu Sasaki, Etsuo Segawa, Akito Suzuki, Kanako Suzuki}

\maketitle

\begin{abstract}
We present some numerical results for nonlinear quantum walks (NLQWs) studied by the authors analytically \cite{MSSSS18DCDS, MSSSS18QIP}. It was shown that if the nonlinearity is weak, then the long time behavior of NLQWs are approximated by linear quantum walks. In this paper, we observe the linear decay of NLQWs for range of nonlinearity wider than studied in \cite{MSSSS18DCDS}. In addition, we treat the strong nonlinear regime and show that the solitonic behavior of solutions appears. There are several kinds of soliton solutions and the dynamics becomes complicated. However, we see that there are some special cases so that we can calculate explicit form of solutions.  
In order to understand the nonlinear dynamics, we systematically study the collision between soliton solutions. We can find a relationship between our model and a nonlinear differential equation. 
\end{abstract}

\section{Introduction}\label{sec:intro}
Quantum walks (QWs), which are quantum analog of classical random walks \cite{ABNVW01, FH10Book, Gudder88Book, Meyer96JSP}, are now attracting much interest because of its connection to various regimes in mathematics, physics and applications such as quantum algorithms \cite{AKR05Proc, Childs09PRL, Portugal13Book} and topological insulators \cite{AO13PRB, CGSVWW16JPA, EKOS17JPA, GNVW12CMP, Kitaev06AP, Kitagawa12QIP, KRBD10PRA}.
Nonlinear quantum walks (NLQWs), which are nonlinear versions of QWs, 
have been recently proposed by several authors \cite{GTB16PRA, LKN15PRA, NPR07PRA} and in particular related to some nonlinear differential equations such as nonlinear Dirac equations \cite{LKN15PRA}.
For experimental realization of QWs by Bose-Einstein condensation, which can realize the nonlinearity in principle, see \cite{AW17PRA,DGGWS18PRL}.
In \cite{MSSSS18DCDS}, we have initiated an analytical study of NLQWs using the methods developed for the study of nonlinear dispersive equations. 
More precisely, it was shown in \cite{MSSSS18DCDS} that the scattering phenomena for NLQWs in a weak nonlinear regime, that is, the behavior of solutions of NLQWs can be approximated by that of corresponding to linear quantum walks. 
We further proved a weak limit theorem, which is one of the main theme of the study of QWs since the celebrated result by Konno \cite{Konno02QIP}, for NLQWs in \cite{MSSSS18QIP}. 

Since the previous papers \cite{MSSSS18DCDS, MSSSS18QIP} discussed the similarity between QWs and NLQWs, in this paper we discuss a characteristic of the nonlinear dynamics, which is a soliton solution. 
The soliton solutions, which are localized traveling waves emerging from the balance of linear dispersion and nonlinearity, are one of the most interesting topics in the study of nonlinear differential (and difference) equations.
In particular, we would like to point out that all the papers proposing NLQWs \cite{GTB16PRA, LKN15PRA, NPR07PRA} are discussing the soliton solutions.
In this paper, we choose a particular nonlinearity (rotation type nonlinearity, see \eqref{s-eq1} below) which supports a soliton solution of simple form and present some theoretical and numerical results of the soliton solution.
We remark that our nonlinearity, although it looks quite natural, is not motivated by physical phenomena.
However, we think the understanding of soliton solutions with a particular nonlinearity will be the first step for the understanding of soliton solutions of more general NLQWs including the models appearing in \cite{GTB16PRA, LKN15PRA, NPR07PRA}.

We now introduce the NLQWs we study in this paper.
Let $C:\R \times \R\to U(2)$, where $U(2)$ is the set of $2\times 2$ unitary matrices. For a matrix $A \in U(2)$ and a vector $u \in \C^2$, $A^T$ and $u^T$ denote the transpose of $A$ and the transpose of $u$, respectively.    
We define the (nonlinear) quantum coin $\hat C: l^2 (\mathbb{Z}; \mathbb{C}^2) \to l^2 (\mathbb{Z}; \mathbb{C}^2)$ by 
\begin{equation}
\label{defC}
(\hat Cu)(x)=C(|u_1(x)|^2,|u_2(x)|^2) u(x),
\end{equation} 
where $u (\cdot) = (u_1 (\cdot) ,u_2 (\cdot) )^T \in  l^2 (\mathbb{Z}; \mathbb{C}^2)$. 
For $(T_\pm u)(x)=u(x\mp 1)$, we set 	
	\[ S=\begin{pmatrix} T_- & 0 \\ 0 & T_+ \end{pmatrix}, \]
that is, 
	\[ (Su)(x)=\begin{bmatrix} \langle e_1, (T_-u)(x) \rangle \\ \langle e_2, (T_+u)(x) \rangle \end{bmatrix}
        = \begin{bmatrix} u_1(x+1)  \\ u_2(x-1) \end{bmatrix}. \]
Here, $e_j \ (j = 1, 2)$ denote $e_1=(1, 0)^T$ and $e_2= (0, 1)^T$. 
Let the one-step nonlinear time evolution operator $U$ depending on a given state $u\in l^2(\mathbb{Z},\mathbb{C}^2)$ be 
\begin{align}
U:=U(u):=S\hat C \label{m-eq2}
\end{align}
which is a map from $ l^2 (\mathbb{Z}; \mathbb{C}^2)$ to itself. 
Then, the state $u^t$ of the walker at time $t$ is defined by the recursion relation 
\begin{align}\label{3}
u^t = U(u^{t-1}) \cdot u^{t-1}, \quad t=1,2,\cdots,
\end{align}
with some initial state $u^0\in l^2(\mathbb Z; \mathbb C^2)$ with $\|u_0\|_{l^2}=1$. 
Notice that $l^2$ norm will be conserved, i.e. $\|u^t\|_{l^2}=1$ for all $t\in \N$. 
We define the nonlinear evolution operator $U(t)$ as 
\begin{align}
U(t)u^0 = u^t. \label{eq-su1}
\end{align}

Throughout this paper, we shall consider the following concrete quantum coin:
\begin{align}
\hat{C} (s_1, s_2) = R\left(\dfrac{\pi}{4}\right)R\left(g(s_1 +  s_2)^p \right), \label{s-eq1}
\end{align}
where $g \in \mathbb{R}$, $p\geq 1$ and $R(\theta)$ is the $\theta$-rotation matrix, that is, 
	\[R(\theta)=\begin{pmatrix} \cos\theta & -\sin\theta \\ \sin \theta & \cos \theta \end{pmatrix}. \]
Thus for given $u$, $\hat{C}$ acts as 
	\[ (\hat{C} u)(x) = R\left(\dfrac{\pi}{4}+g||u(x)||^{2p}_{\mathbb C^2}\right)u(x), \]
where $\|(u_1,u_2)^T\|_{\mathbb C^2}^2=|u_1|^2+|u_2|^2$.
The value of $|g|$ is a strength of the nonlinearity. 
Thus in the case where $|g|$ is not small, it is expected that a nonlinear effect arises.  
Indeed, in \cite{MSSSS18DCDS}, it has been shown that $u^t$ scatters for $p \ge 2$ and $|g|\ll1 $.
To this end, let us prepare the following remark. 
In order to make the dependence on $g$ explicit, we write $U_g (t)$ for $U(t)$ defined by \eqref{eq-su1} with the coin \eqref{s-eq1}. 
We observe that, for $v^{0} = |g|^{p/2} u^{0}$ with $\|u^{0}\|_{l^2}=1$, 
\[
U_g (t) u^{0} = \frac{1}{\sqrt{|g|}} U_{g=1}(t) v^{0}
\]
since $g||u^{0}(x)||^{2p}_{\mathbb C^2} + \pi/4=||v^{0}(x)||^{2p}_{\mathbb C^2} + \pi/4$. 
Hence, we can fix $|g|=1$ and vary the norm value of $\|u_0\|_{l^2}$. 
Here, a small $\|u_0\|_{l^2}$  corresponds to the case where $|g|$ is small which means a small non-linear strength and vice versa.  
Therefore, we study the cases $g = \pm 1$ 
for \eqref{s-eq1} and see the dependence on the initial states. 

The organization of this paper is the following.
In Section \ref{sec:solitonic}, we give some explicit examples of soliton solutions of NLQWs and discuss their stability/instability.
In Section \ref{sec:linfty}, we numerically study the behavior of $l^\infty$ norm ($\|u\|_{\l^\infty}=\sup_{x\in\Z}\|u(x)\|_{\C^2}$) and observe three types of behavior, which are linear decay, soliton and oscillation.
The oscillation solution is a new type of solution and we will give some theoretical explanation in Subsection \ref{subsec:osci}.
In Section \ref{sec:coll}, we systematically study the collision of solitons by numerical simulations.
In Section \ref{sec:disc}, we summarize our results.

\section{Decision of the solitonic behavior in a dynamical system}\label{sec:solitonic}
\begin{figure}[t]
\begin{center}
\includegraphics[width=12.0cm]{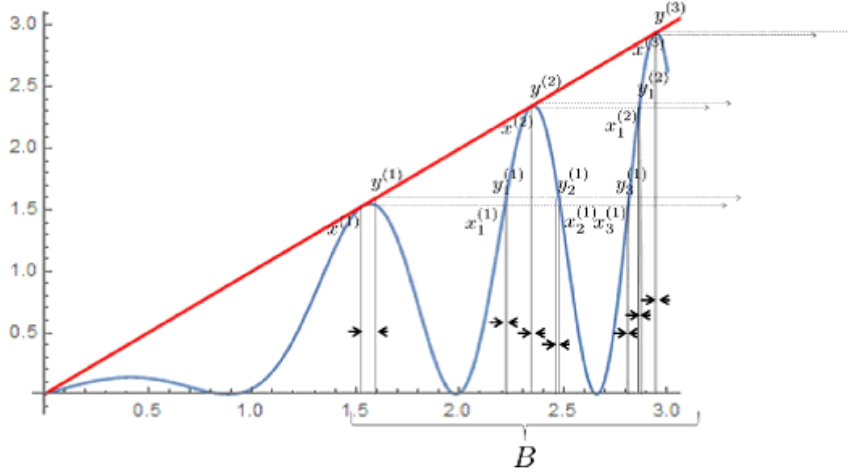} 
\caption{The set of intial data giving the solitonic behavior $B$: The curve depicts $y=f(x)$ and the diagonal line depicts $y=x$.}
\label{fig_solitonset}
\end{center}
\end{figure}
In this section, we discuss the existence and stability of soliton solutions of NLQWs.
First, we give an example of explicit non-scattering solutions. 
More precisely, we let $j = 1, 2$ and define $\delta_{j,x}\in l^2 (\Z;\C^2)$ as $\delta_{j,x}(y)= e_j$ if $y=x$ and $\delta_{j,x}(y)=0$ if $y\neq x$. 
Then, setting
\begin{align}
\varphi^t =a \delta_{1,-t}, \label{eq:soliton}
\end{align}
where $a$ is a root of $\pi / 4 + g a^{2p}=0$, $\varphi^t$ becomes a solution of NLQWs. 
It is observed that this solution is localized, travels by constant speed and does not decay.
Thus, we call this solution a soliton solution.
Further, observe that if we take initial data having several solitons, that is taking $u^0=\sum_{j=1}^N a \delta_{1,x_j}$ with $x_j\neq x_k$ for $j\neq k$ and $a$ as above, we see that each soliton does not interact with each other and we will have a multi-soliton solution:
\begin{align}
\varphi^t =\sum_{j=1}^Na \delta_{1,x_j-t}. \label{eq:multisoliton}
\end{align}
We remark that although we have a family of multi-soliton solutions as the usual KdV or cubic Schr\"odinger equation, the situation is a little different because the only explicit solitons which we can construct are with constant velocity $\pm1$ or the one appearing in Remark \ref{rem1} below with velocity essentially $0$.
In spite of these differences, we will observe typical phenomena for soliton solutions, which behave like KdV solitons in Section \ref{sec:coll}. 
That is, when two solitons collide, they pass though each other without changing their shapes but only changing their positions and phases. 

\begin{remark}\label{rem1}
It is also possible to have an extreme opposite behavior to the solitonic behavior as a nonlinear effect adjusting a parameter: 
if $\pi / 4 + g a^{2p}=\pi/2$, then there is a periodic solution localized in a finite region 
with period $4$, which evolves a zig-zag walking; 
$a \delta_{1,0}\to a \delta_{2,1}\to -a \delta_{1,0}\to -a \delta_{2,1} \to a \delta_{1,0}\to \cdots$. 
\end{remark}

It can be shown that 
for the existence of such non-scattering solutions, it suffices to have $\pi /4 + g a^{2p}\in \pi n \ (n \in \mathbb{N})$ 
so even for $g>0$. Then, there exist solutions with solitonic behavior. 
Moreover we can show that $\varphi^t$ defined by \eqref{eq:soliton} is unstable.
Indeed, for $ 0 < \varepsilon < 1$, 
we let $\varphi_\varepsilon (t)$ be a solution of NLQW with its initial state $\varphi_{\varepsilon}(0):=a(1-\varepsilon)\delta_{1,0}$. 
Then, it is clear that $\varphi_\varepsilon (0) \to \varphi^0$ as $\varepsilon\downarrow 0$. 
We see that $\|\varphi_\varepsilon^t (-t)\|_{\C^2}\to 0$ as $t \to \infty$ 
and obtain $\|\varphi^t - \varphi_\varepsilon^t\|_{l^2}\to \|\varphi^t\|_{l^2}+\|\varphi_\varepsilon^t\|_{l^2}\sim 1$ as $t \to \infty$.
On the other hand, if a solution of NLQW $\psi_\varepsilon^t$ starts from the following state
\begin{align*}
\psi_\varepsilon^0 (x)=\begin{cases} 0 & x<0\\ (1+\varepsilon)\begin{pmatrix} a \\ 0 \end{pmatrix} & x=0\\
\text{whatever} & x>0,
\end{cases}
\end{align*}
then 
we see that $\psi_\varepsilon^t (-t)\to (a, 0)^T$ as $t \to \infty$. 
Thus, there is a large set of initial data which persists soliton-like behavior with speed $1$  in the left edge. 
We will explain this picture for the soliton using a dynamical system as follows.

\begin{figure}[t]
\begin{tabular}{cc}
\begin{minipage}{0.45\hsize}
(a)
\vspace{-0.5cm}
\begin{center}
\includegraphics[width=6.0cm]{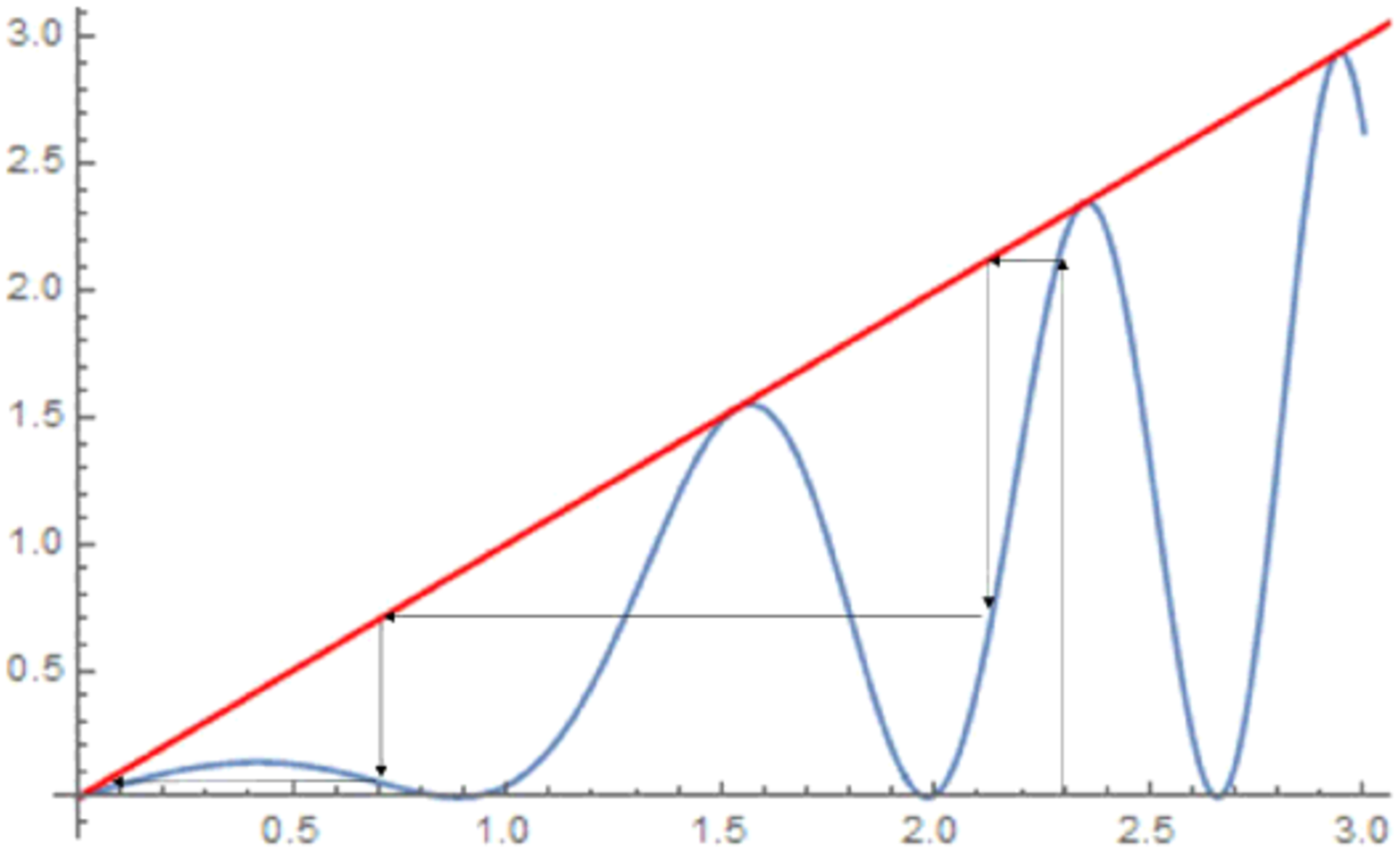} 
\end{center}
\end{minipage}
\hspace{6mm}
\begin{minipage}{0.45\hsize}
(b)
\vspace{-0.5cm}
\begin{center}
\includegraphics[width=6.0cm]{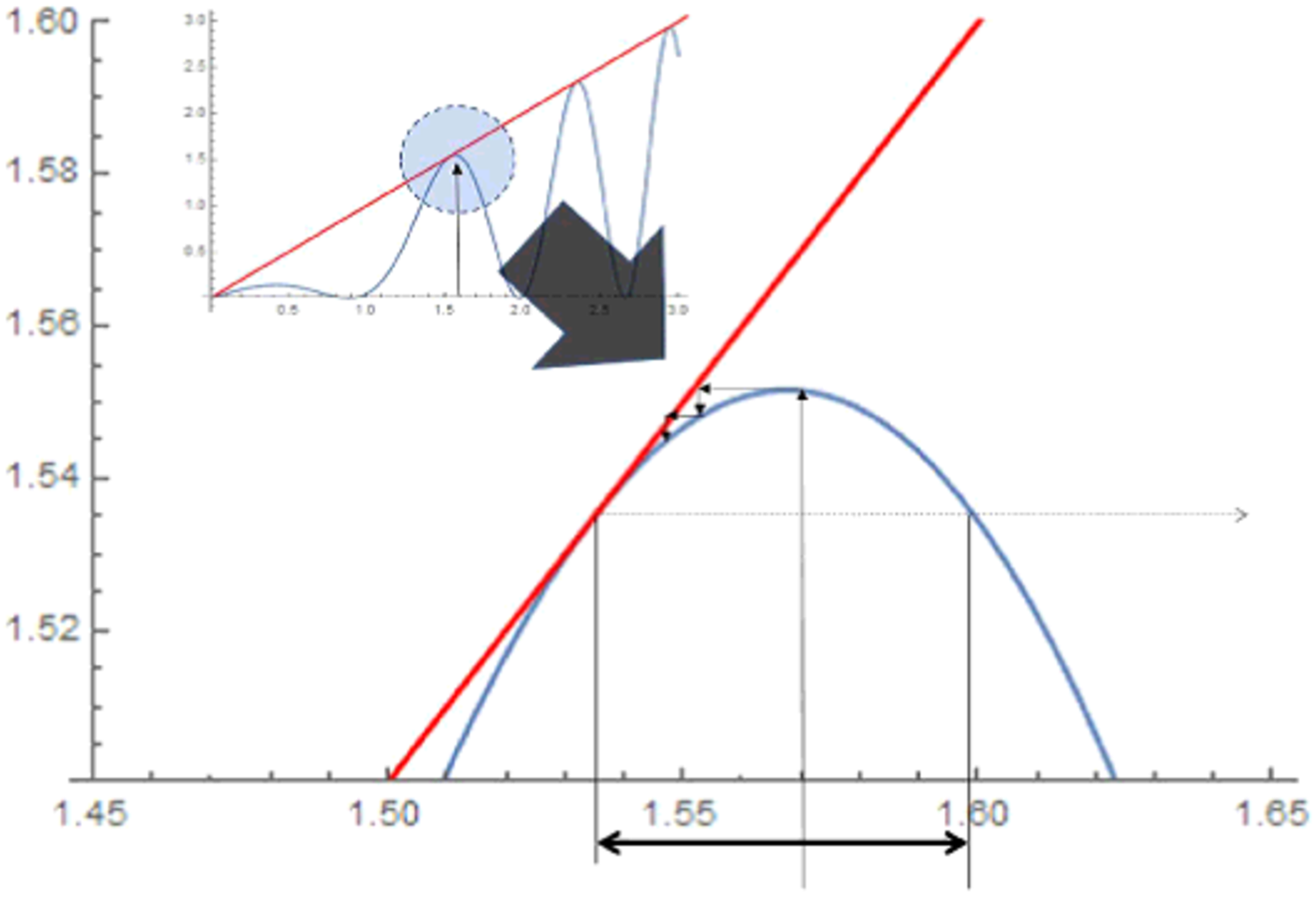} 
\end{center}
\end{minipage}
\end{tabular}
\caption{The dependence on the initial data: 
Figure (a) depicts the fixed point is $0$ since the initial data $u_0\notin B$.  
On the other hand, in (b), we see that the fixed point is a strictly positive value belonging to $P$. }
\label{fig_fixpoint}
\end{figure}

Our interest is now the left most amplitude at each time step, that is, $u^{t}(-t)$. 
Then we regard the initial state as $u^{0}(x)=\delta_0(x)e_1$. 
We put $r_t:=||u^{t}(-t)||^2_{l^2}$ and $\theta_t:=\pi/4+gr_t^p$. Here $g\in \{\pm 1\}$. 
Since $u^{t}(-t)=\cos \theta_{t-1}\cdot u^{t-1}(-t+1)$, we have 
	\begin{equation}\label{eq:dynamical}  
        r_{t+1}=r_t \cos^2[\pi/4 + gr_t^p]. 
        \end{equation}
Then the problem is simply reduced to the dependence on the initial data $r_0$ 
with respect to the convergence destination in the above dynamical system (\ref{eq:dynamical}). 
Now to obtain the fixed point of this discrete-time dynamical system, 
let us consider the following function. 
	\begin{equation}
        f(x)=x\cos^2 (\pi/4+gx^p),\;\;(x\geq 0).
        \end{equation}
We pick up important properties of $f(x)$ as follows. 
\begin{enumerate}
\renewcommand{\labelenumi}{(\roman{enumi})}
\item $f(x)=x$ if and only if $x=0$ or $gx^p\in (4\mathbb{Z}+3)\pi/4$;
\item $f(x)$ takes a local maximum if and only if $\cot[\pi/4+gx^p]=2pgx^p$; 
\item $f(x)$ takes a local minimum\footnote{Every local minimum value is $0$.} if and only if $gx^p\in (4\mathbb{Z}+1)\pi/4$;
\item $f(x)\leq x$ ($x\geq 0$). 
\end{enumerate}
Remark that the points which accomplish (i), (ii), (iii), respectively, appear periodically as follows: 
${\rm (i)} \to {\rm (ii)} \to {\rm (iii)} \to {\rm (i)} \to {\rm (ii)} \to {\rm (iii)} \to {\rm (i)} \to\cdots$ which is independent of $g$ and $p$. 
We set 
$P:=\{x\geq 0 \;|\; gx^p\in (4\mathbb{Z}+3)\pi/4\}$, 
which is the set of all the points satisfying (i) except $0$. 
Let $x^{(m)}$ be the $m$-th smallest element of $P$. 
Let $y^{(m)}>x^{(m)}$ be the smallest value of the solution of $x^{(m)}=f(x)$. 
We define $P^{(m)}:=\{x\in \mathbb{R}_{+}\setminus\{y_m\} \;|\; x^{(m)}=f(x)\}$, and 
$Q^{(m)}:=\{x\in \mathbb{R}_{+} \;|\; y^{(m)}=f(x)\}$.
We put $P^{(m)}=\{x_1^{(m)},x_2^{(m)},\dots\}$ and $Q^{(m)}=\{y_1^{(m)},y_2^{(m)},\dots\}$ with 
$x_n^{(m)}<x_{n+1}^{(m)}$, $y_n^{(m)}<y_{n+1}^{(m)}$, see Fig.~\ref{fig_solitonset}.
Then from a simple observation to this iterative dynamical system (see Figs.\ref{fig_fixpoint}),  
we can completely determine the set in $\mathbb{R}_+$ for the initial value $r_0$ providing a solitonic behavior whose fixed point is $x^{(m)}$:
\[ B^{(m)}:= [x^{(m)},y^{(m)}]\cup [x^{(m)}_1,y^{(m)}_1] \cup [y^{(m)}_2,x^{(m)}_2]\cup[x^{(m)}_3,y^{(m)}_3] \cup [y^{(m)}_4,x^{(m)}_4]\cdots  . \]
Then we have 
	\begin{equation}
        \lim_{n\to\infty}r_n>0 \Leftrightarrow r_0\in \bigcup_{m\geq 0}B^{(m)}=:B. 
        \end{equation}
Therefore for small value $\epsilon>0$, if $r_0=x^{(m)}-\epsilon$, then the fixed point of the dynamical system is $0$, 
while if $r_0=x^{(m)}+\epsilon$, then the fixed point is $x^{(m)}(>0)$. 

\begin{remark}
We do not exclude the possibility that $\varphi_\varepsilon$ converges (in some sense) to another unknown traveling wave type solution.
\end{remark}

\section{Behavior of $l^\infty$ norm}\label{sec:linfty}
\begin{figure}[tbhp]
\begin{tabular}{cc}
\begin{minipage}{0.45\hsize}
(a)
\begin{center}
\includegraphics[width=6.0cm]{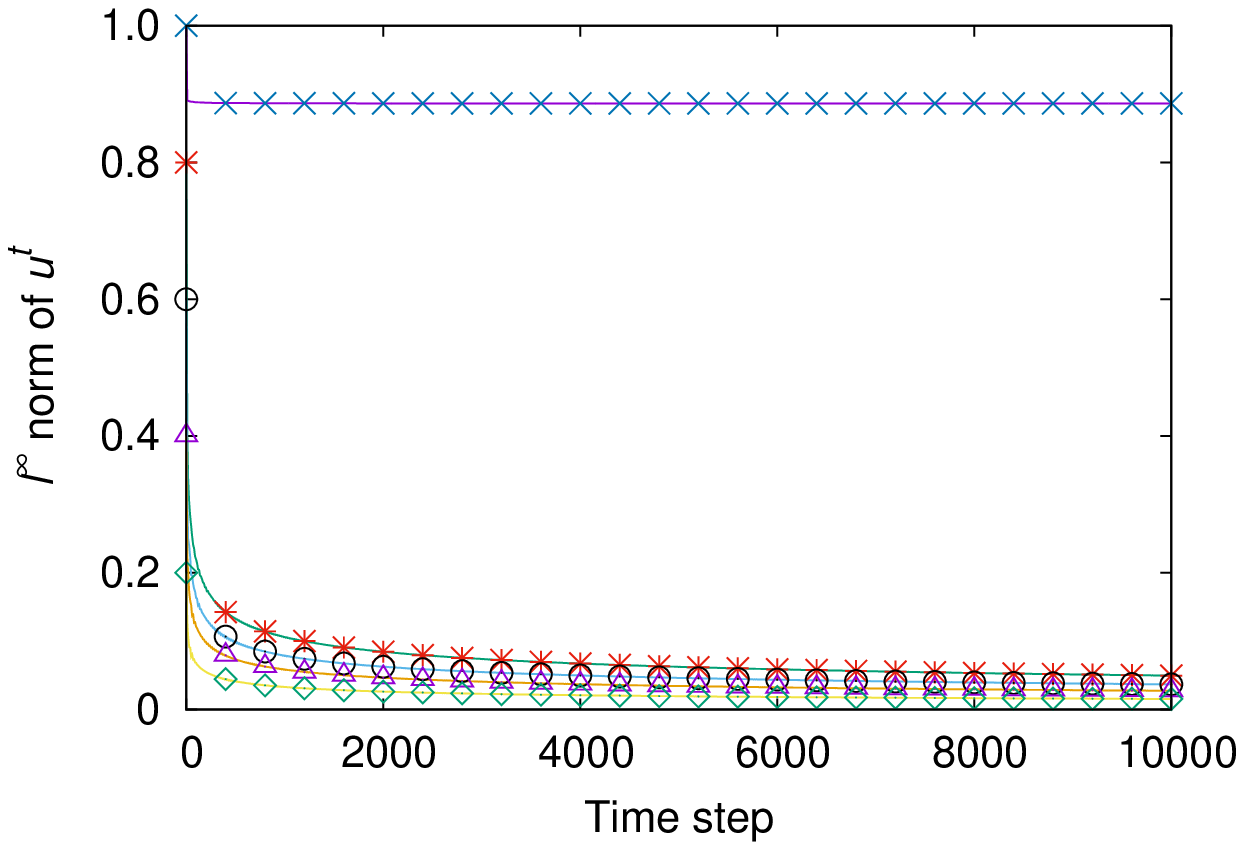} 
\end{center}
\end{minipage}
\hspace{6mm}
\begin{minipage}{0.45\hsize}
(b)
\begin{center}
\includegraphics[width=6.0cm]{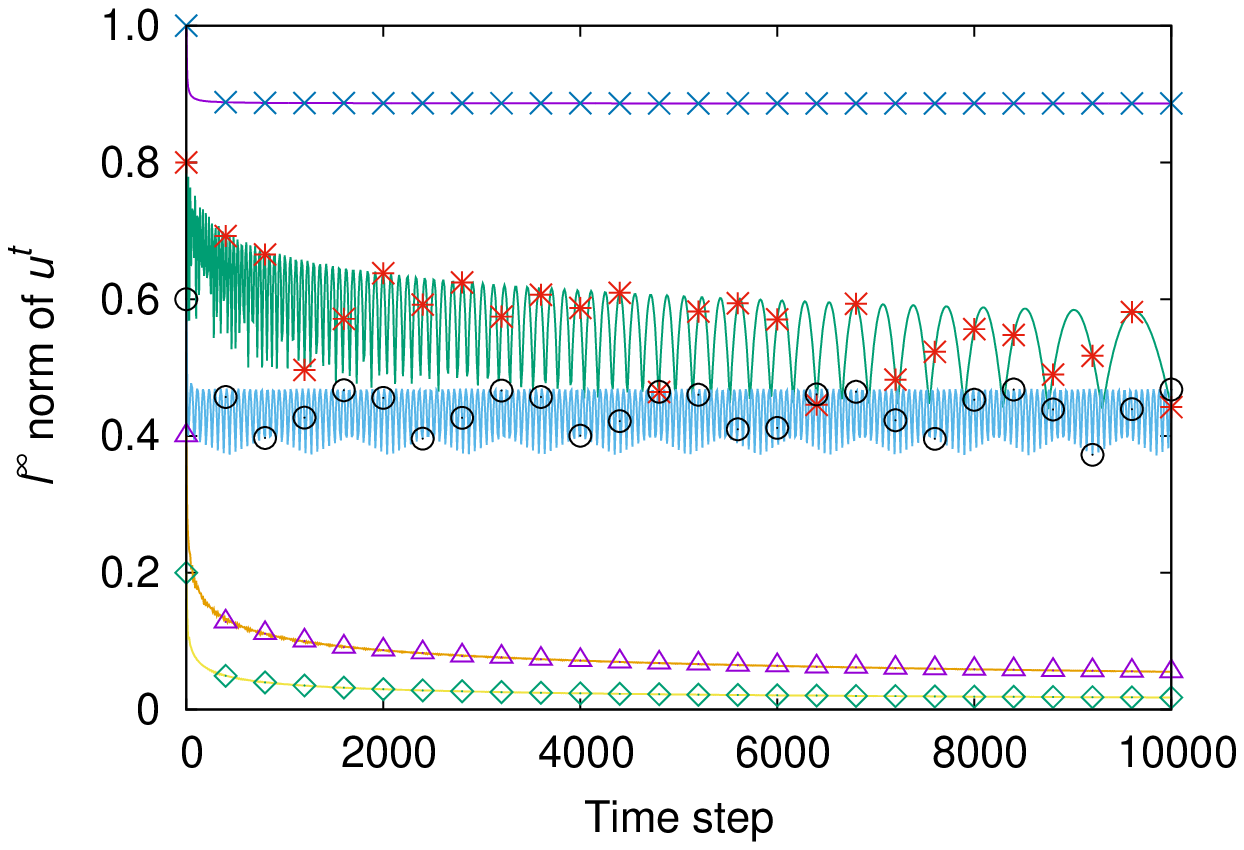} 
\end{center}
\end{minipage}
\end{tabular}
\caption{Behaviors of $\|u^t \|_{l^\infty}$ to the model \eqref{s-eq1}. 
Figure (a) is obtained by coin $C_+$, and Figure (b) is obtained by coin $C_-$. In both figures, each curve starts $u^0 = \delta_{1, 0}\ (\times)$, $u^0 = 0.8 \delta_{1, 0}\ (\ast)$, $u^0 = 0.6\delta_{1, 0} \ (\circ)$, $u^0 = 0.4\delta_{1, 0} \ (\triangle)$, $u^0 = 0.2\delta_{1, 0}\ (\diamond)$.  
In (b), we see oscillations of $\|u^t \|_{l^\infty}$ for the cases $u^0 = 0.8\delta_{1, 0}$ and $u^0 = 0.6 \delta_{1, 0}$, which do not appear in (a). }
\label{sup-norm-u}
\end{figure}

In this section, we study the behavior of $l^\infty$ norm of the solutions.
We set $C_\pm$ the coin \eqref{s-eq1} with $g = \pm1$ and $p = 1$.

\begin{figure}[tbhp]
\begin{tabular}{cc}
\begin{minipage}{0.45\hsize}
(a)
\begin{center}
\includegraphics[width=6.0cm]{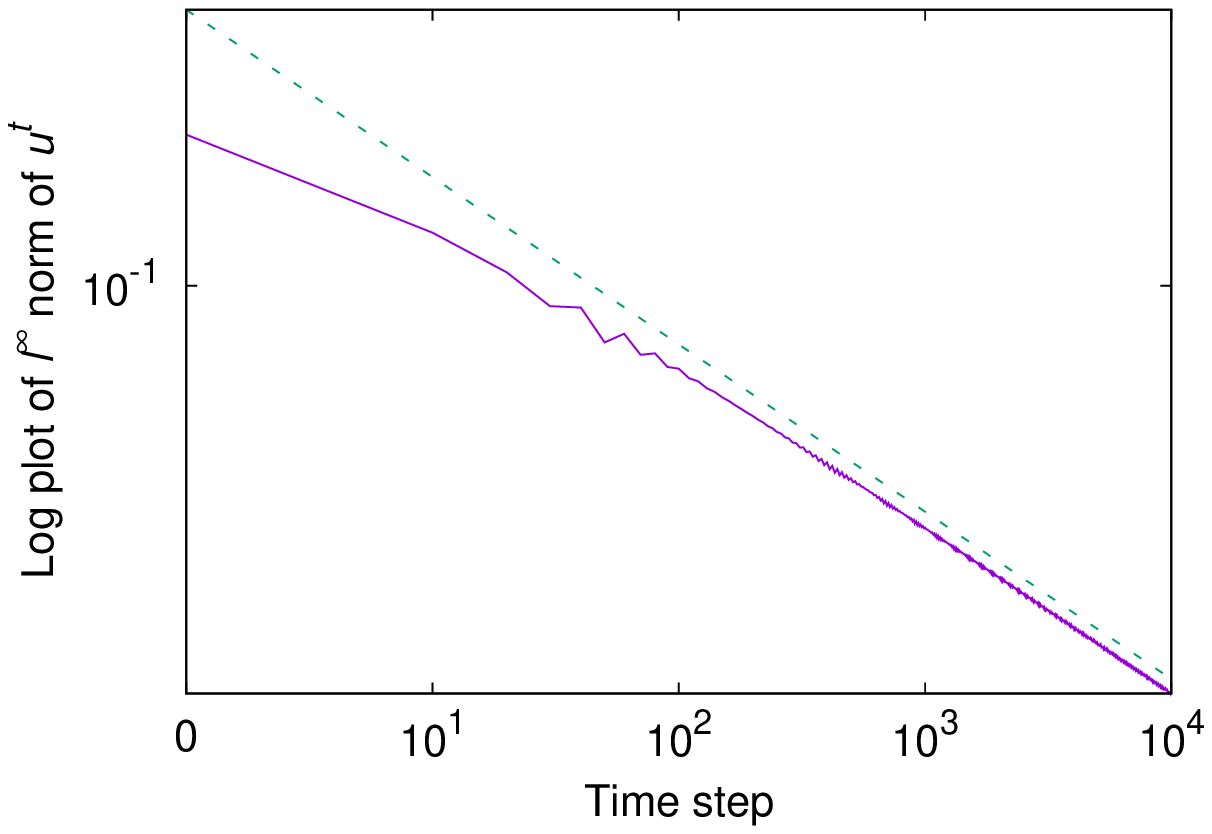}  
\end{center}
\end{minipage}
\hspace{6mm}
\begin{minipage}{0.45\hsize}
(b)
\begin{center}
\includegraphics[width=6.0cm]{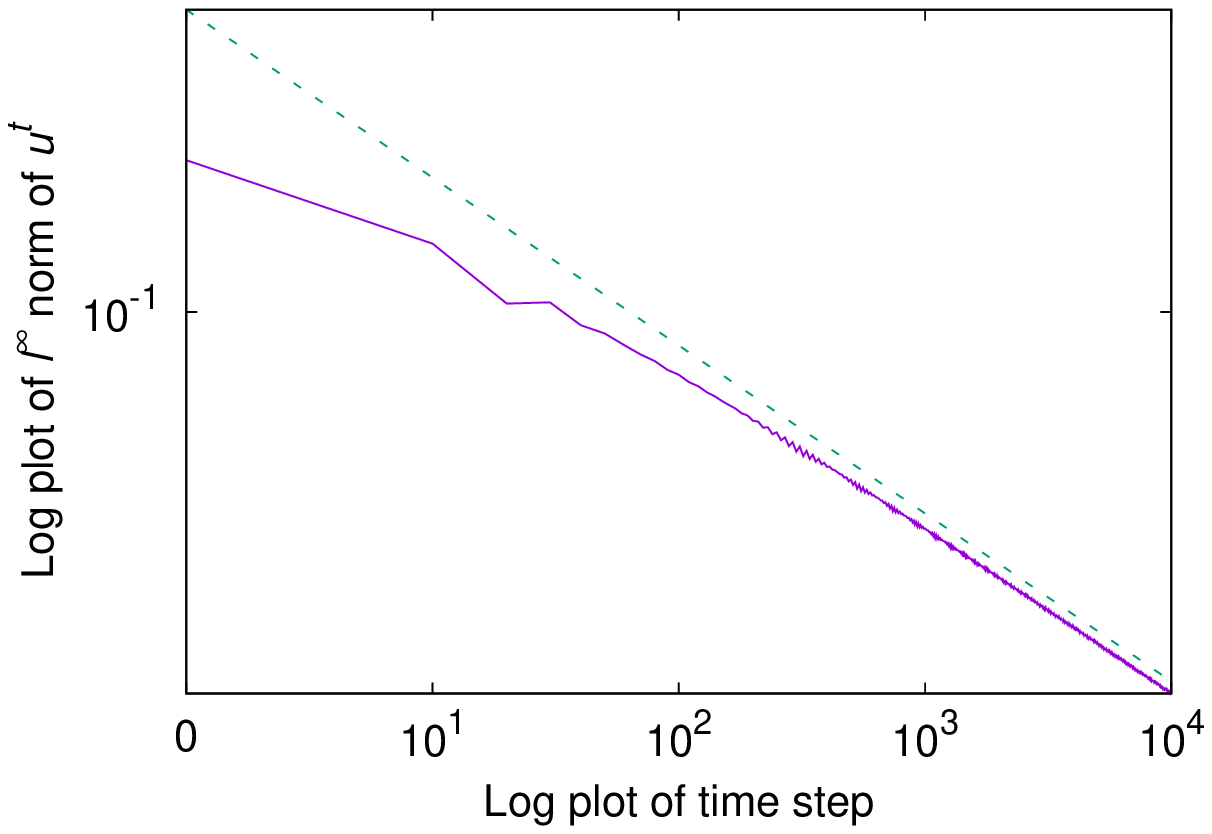} 
\end{center}
\end{minipage}
\end{tabular}
\caption{Log-log plot of the behavior of $\|u^t \|_{l^\infty}$ with $u^0 = 0.2 \delta_{1, 0}$ to the model \eqref{s-eq1}, where (a) is for the coin $C_+$, and (b) is for $C_-$. The broken lines are given by $-t/3.0 - 0.40$ in (a) and $-t/3.0 -0.45$ in (b), which are good approximations. }
\label{log-log}
\end{figure}

First, we discuss the behavior of $\|u^t\|_{l^\infty}$ with small initial state, which implies that the nonlinear effect is week. 
From Theorem 2.1 of \cite{MSSSS18DCDS}, the behavior of $\|u^t \|_{l^\infty}$ of the linear model (the case $g = 0$ in \eqref{s-eq1}) is approximated by $t^{-1/3}$ as $t \to +\infty$. It can be verified that $\|u^t \|_{l^\infty}$ to the nonlinear model \eqref{s-eq1} with small initial state behaves like a linear.  Indeed, Figure \ref{log-log} shows that the behavior with $u^0 = 0.2\delta_{1, 0}$ is approximated by a linear model.
We remark that the case $p=1$ was not proved to scatter in \cite{MSSSS18DCDS}.

On the other hand, if $u^0 = \delta_{1, 0}$ for both coins $C_\pm$, then $\|u^t\|_{l^\infty}$ does not decay and becomes almost constant (see Figure \ref{sup-norm-u}). Thus, it is expected that the solitonic behavior which has been discussed in Section 2 appears. For this case, we can find the explicit value of $a = \|u^t\|_{l^\infty}$ by solving the equation $\pi/4 + a^2 = \pi/2$ for $C_+$ and $\pi/4 -a^2 = 0$ for $C_-$.  
In Table 1, we put the value of $\|u^t \|_{l^\infty}$ obtained by the numerical calculations starting from $u^0 = \delta_{1, 0}$, which can be compared with $a = \left(\pi/4\right)^{1/2} \approx 0.886227$. 
\begin{table}[h!]\label{table1}
\begin{center}
\begin{tabular}{|c|c|c|c|}
\hline
 & Numerical value of $\|u^t\|_{l^\infty}$ with $u^0 = \delta_{1, 0}$ \\
\hline
$C_+$ &  0.886256\\
\hline
$C_-$ &  0.886299\\
\hline

\hline
\end{tabular}
\caption{Numerical results are obtained at $t = 10000$. }
\end{center}
\end{table}

\begin{figure}[tbhp]
\begin{center}
\includegraphics[width=6.5cm]{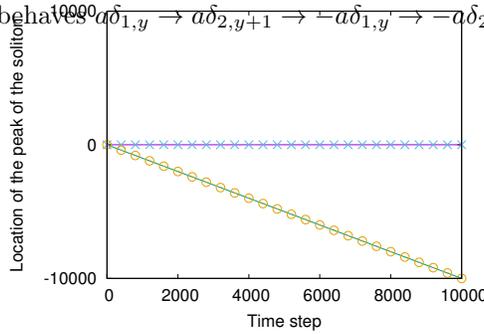}  
\caption{Location of the peak of the soliton with $u^0 = 0.886227\delta_{1, 0}$. The lines with $'\times'$ and with $'\circ'$ correspond to the case $\pi/4 + a^2 = \pi/2$ and $\pi/4 -a^2 = 0$, respectively. }
\label{fig:soliton}
\end{center}
\end{figure}

It is clear that a solution corresponding to the equation  
\begin{align}
\dfrac{\pi}{4} - a^2 = 0 \label{eq-Tsoliton-}
\end{align}
is soliton solution which has been discussed in Section 2. Hence, it is expressed by $u^t = a \delta_{1, -t}$ when $u^0 = a\delta_{1, 0}$, which keeps its shape and moves to left on the $x$-axis with its speed $1$ (see Figure \ref{fig:soliton}). 
As it has been already mentioned in Remark \ref{rem1}, solving the equation
\begin{align}
\dfrac{\pi}{4} + a^2 = \dfrac{\pi}{2}, \label{eq-Psoliton+}
\end{align}
we get a periodic solution, which behaves $a\delta_{1, y} \to a \delta_{2, y+1} \to -a \delta_{1, y} \to -a \delta_{2, y+1} \to a \delta_{1, y} \to \cdots$. 

\subsection{Oscillations}\label{subsec:osci}
A striking feature of the coin $C_-$ is an oscillating behavior of $\|u^t\|_{l^\infty}$. In Figure \ref{sup-norm-u} (b), we can see oscillations when $u^0 = 0.8\delta_{1, 0}$ and $u^0 = 0.6\delta_{1, 0}$. These are different from solitons. We would like to consider the reason why such behavior appears.

\begin{figure}[tbhp]
\begin{tabular}{lll}
(a) & (b) & (c) \\
\includegraphics[width=4.3cm]{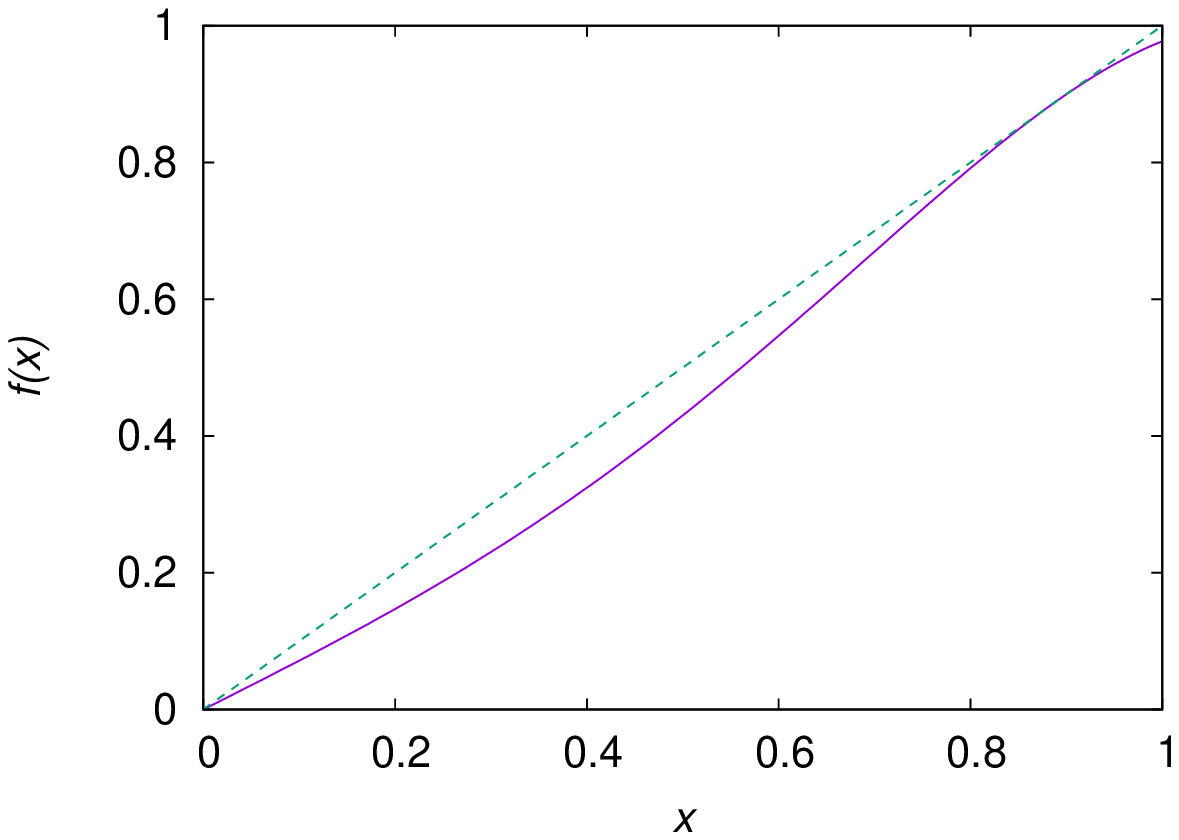}  &
\includegraphics[width=4.3cm]{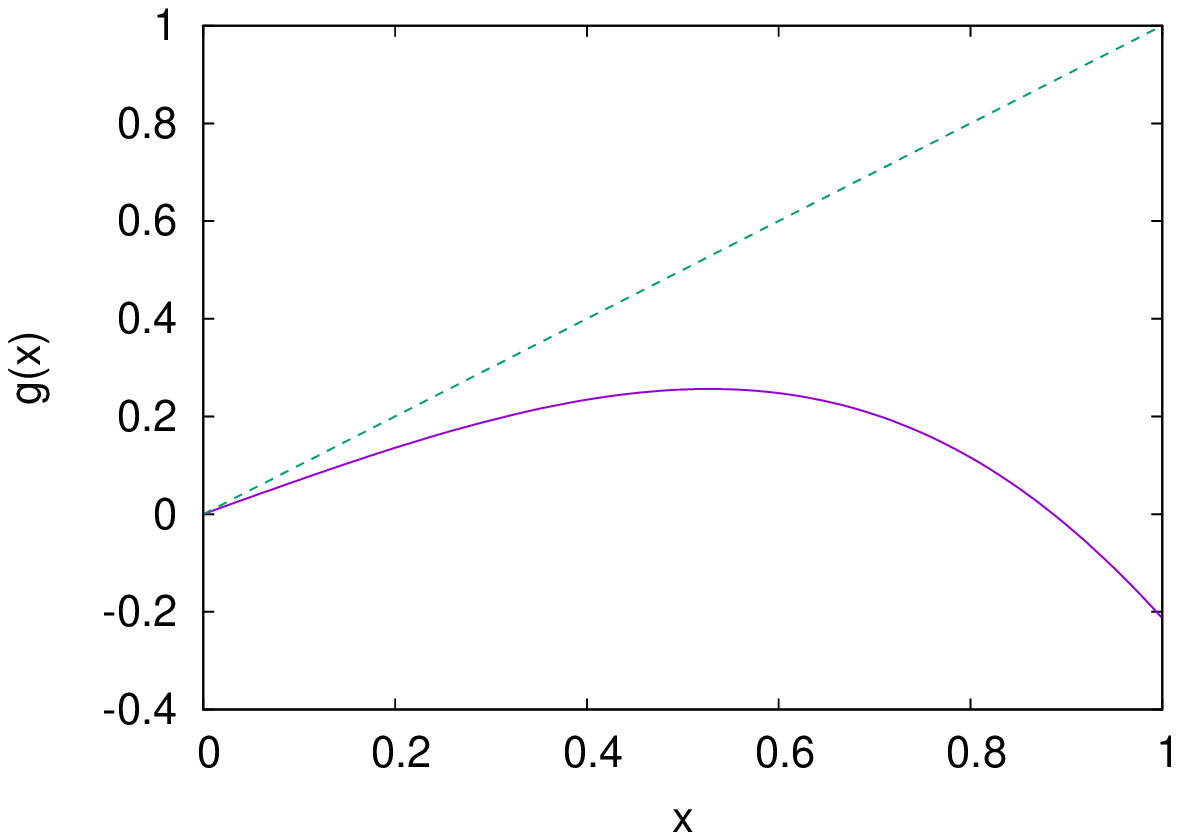}  &
\includegraphics[width=4.3cm]{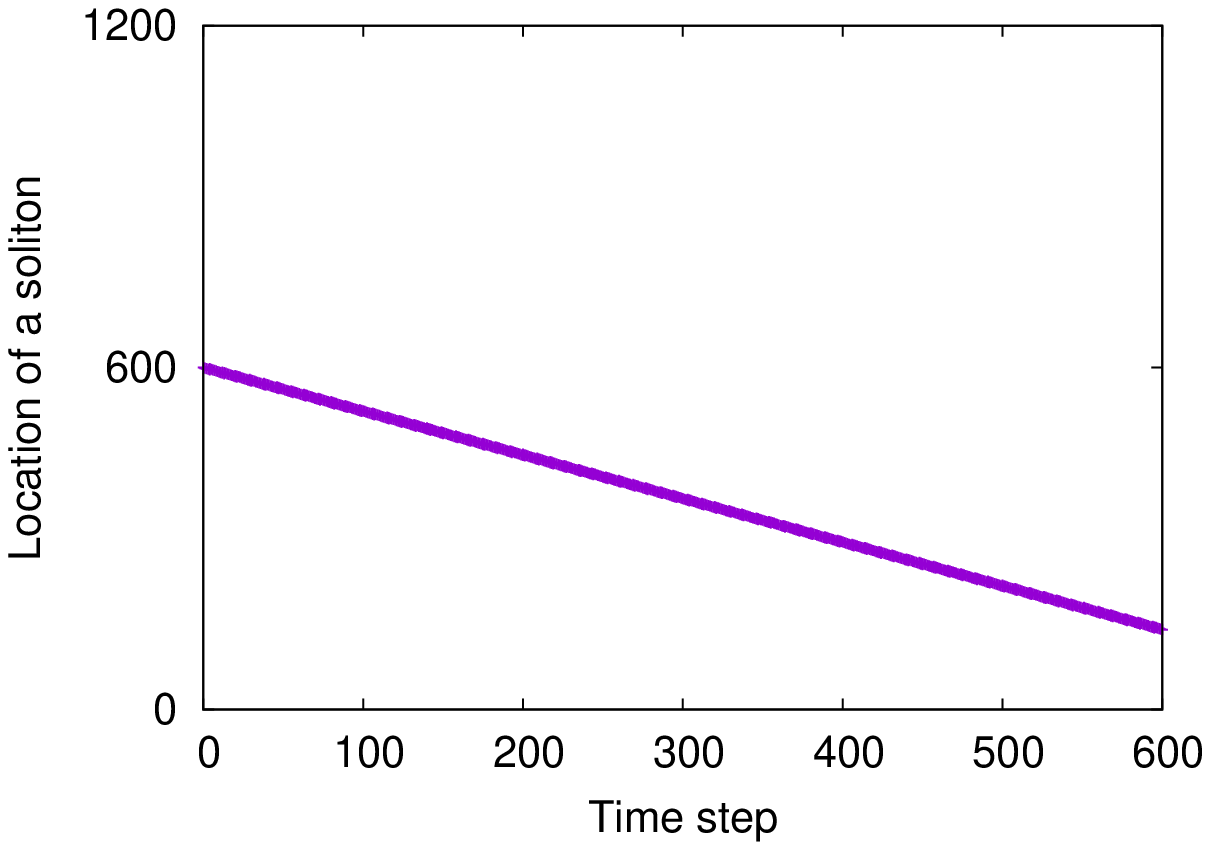}  
\end{tabular}
\caption{(a) A graph of $y = h^- (x)$ with line and a dashed line is $y = x$. (b) A graph of $y = h^+ (x)$ with line and a dashed line is $y=x$. (c) For $u^t$ with $u^0 = 0.6\delta_{1, 600}$ in the case $C_-$, the location of points with $u_1 \ge 0.3$ or $u_2 \ge 0.3$ are plotted.}
\label{fig_oscillating}
\end{figure}

Let $\alpha > 0$ and $u^0 = \alpha \delta_{1, 0}$. Then, it is clear that we have that
\begin{align*}
u^1 &= \alpha \cos{\left(\dfrac{\pi}{4}-\alpha^2\right)}\delta_{1, -1} + \alpha \sin{\left(\dfrac{\pi }{4}-\alpha^2\right)}\delta_{2, 1}, \\
u^2 & = \alpha \cos{\left(\dfrac{\pi}{4}-\alpha^2\right)}\cos{\left(\dfrac{\pi}{4}-\alpha^2 \cos^2{\left(\dfrac{\pi}{4}-\alpha^2\right)}\right)}\delta_{1, -2} + \alpha \cos{\left(\dfrac{\pi}{4}-\alpha^2 \right)} \sin{\left(\dfrac{\pi}{4}-\alpha^2 \cos^2{\left(\dfrac{\pi}{4}-\alpha^2 \right)}\right) }\delta_{2, 0} \\
&\quad - \alpha \sin{\left(\dfrac{\pi}{4}-\alpha^2\right)}\sin{\left(\dfrac{\pi}{4}-\alpha^2 \sin^2{\left(\dfrac{\pi}{4}-\alpha^2\right)}\right)}\delta_{1, 0} + \alpha \sin{\left(\dfrac{\pi}{4}-\alpha^2 \right)} \cos{\left(\dfrac{\pi}{4}-\alpha^2 \sin^2{\left(\dfrac{\pi}{4}-\alpha^2 \right)}\right) }\delta_{2, 2}, \\
& \cdots
\end{align*}
The left and right end parts of $u^t$ are on the function 
\begin{align}
h^- (x) = x \cos{\left(\dfrac{\pi}{4} -x^2 \right)}. \label{0320-eq1}
\end{align}
Since $h^- (0) = 0$ and a graph of $y = h^- (x)$ touches a graph of $y = x$ at $x = \sqrt{\pi}/2$ which is the value in Table 1, a sequence $\left\{x_n \cos{\left(\dfrac{\pi}{4}-x_n^2 \right)}\right\}_{n=2}^\infty$ converges to $0$ if the initial condition is smaller than $\sqrt{\pi}/2 \approx 0.886227$. Thus, oscillation behavior comes from an inner part of $u^t$. Actually, a speed of the main part of $u^t$ is less than $1$ (see Figure \ref{fig_oscillating} (c)). 

The reason why such oscillating solution appears only in the case $C_-$ may be related to the graphs of \eqref{0320-eq1} and \eqref{eq1-0727}. Let the initial state $u^0 = 0.6\delta_{1, 0}$. In the case of $C_+$, it is easily seen that the left and right end parts of $u^t$ are on the function
\begin{align}
h^+ (x) = x \cos{\left(\dfrac{\pi}{4} + x^2\right)}, \label{eq1-0727}
\end{align}
of which graph is in Figure \ref{fig_oscillating} (b). From Figure \ref{fig_oscillating} (b), the decay of $\|u^t\|_{l^\infty}$ is quick and it can be less than $0.3$ after the first step. This decay is also observed in Figure \ref{sup-norm-u} (a). On the other hand, we see from Figure \ref{fig_oscillating} (a) that the left and the right end parts of $u^t$ decay slowly. This implies that the inner part of $u^t$, which denotes the area between the left and the right end parts, can have strong influence coming from the end parts for long time steps. Then, the dynamics becomes complicated. 


\section{Collision between two solitons}\label{sec:coll}
\begin{figure}[tbhp]
\begin{tabular}{cc}
\begin{minipage}{0.45\hsize}
(a)
\begin{center}
\includegraphics[width=6.0cm]{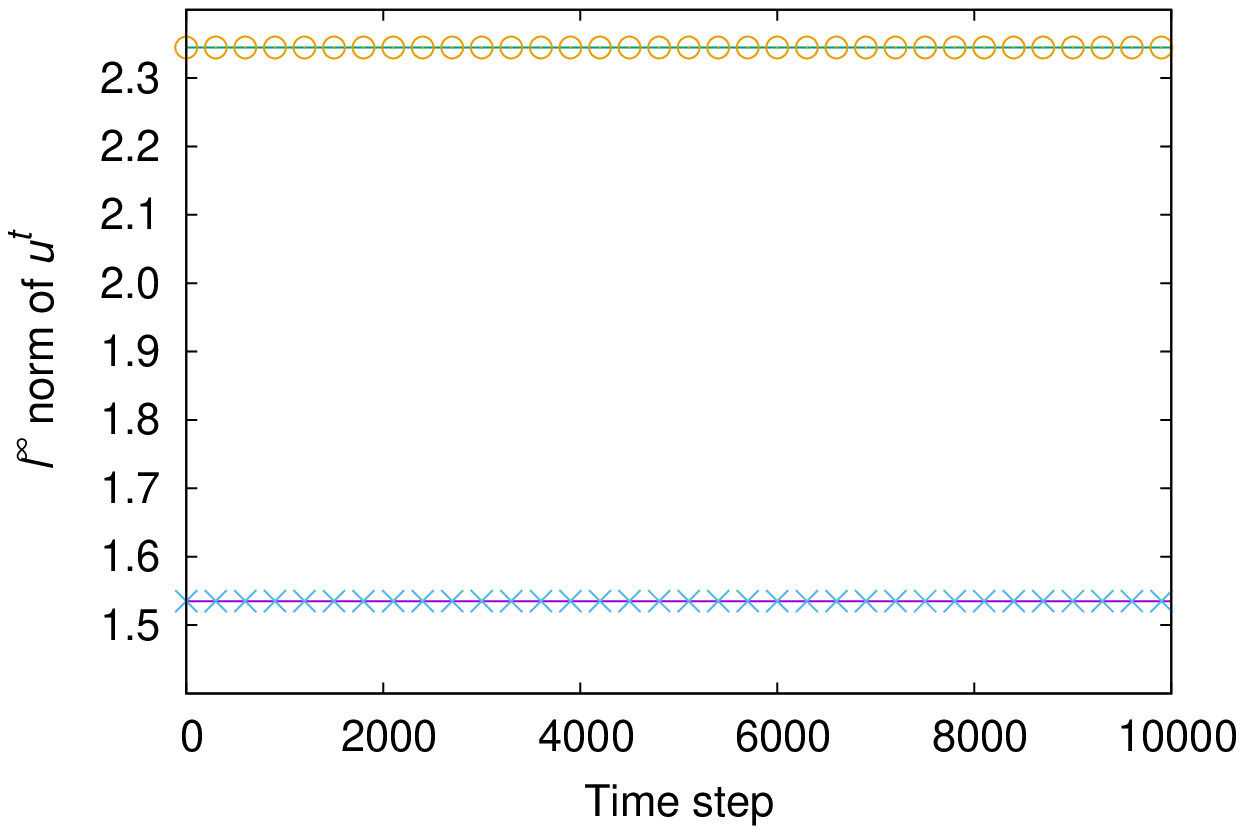}  
\end{center}
\end{minipage}
\hspace{6mm}
\begin{minipage}{0.45\hsize}
(b)
\begin{center}
\includegraphics[width=6.0cm]{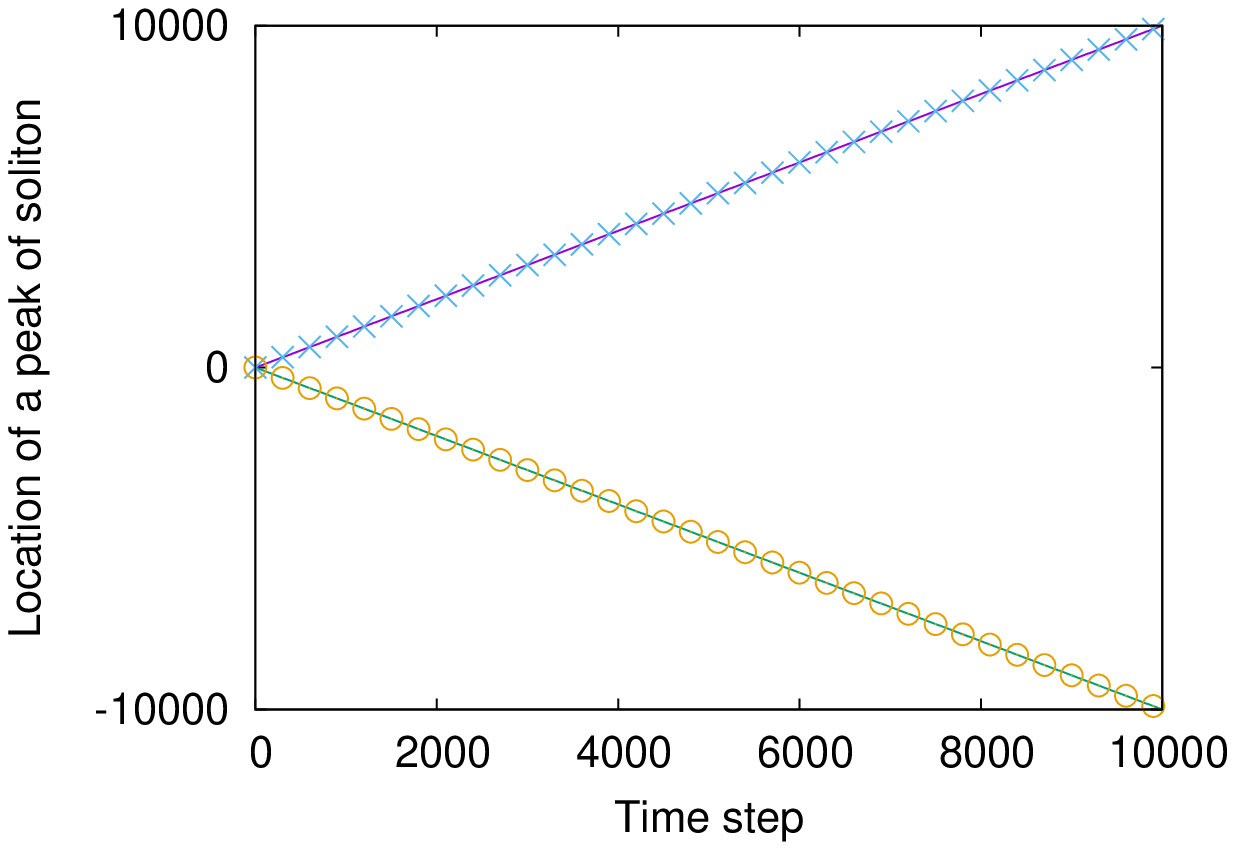} 
\end{center}
\end{minipage}
\end{tabular}
\caption{Solitons for the coin $C_+$ with $u^0 = 1.534990\delta_{2,0}$ ($\times$) and with $u^0 = 2.344736\delta_{1, 0}$ ($\circ$). (a) is the  behavior of $\|u^t\|_{l^\infty}$ from $t = 0$ to $10000$. (b) shows the location of each soliton starting from the origin. The soliton with $u^0 = 1.534990\delta_{2, 0}$ moves to the right and the one with $u^0 = 2.344736\delta_{1, 0}$ moves to the left on the $x$-axis.}
\label{soliton+RT}
\end{figure}
If we take $\|u^0\|_{l^2} > 1$, which implies a nonlinear effect becomes stronger, then several kinds of non-scattering solutions appear. Among such solutions, we focus on solitons.  Considering a nonlinear system, we are interested in not only the existence of solitons but also an interaction between solitons. We would like to consider a collision between two solitons and observe the behavior of solitons after collision. 

We see that soliton solutions from Section 2 are divided into two types. Indeed, 
for the case with coin $C_+$, solve a equation
\begin{align}
\dfrac{\pi}{4} + a^2 = (2n-1)\pi \qquad (n = 1, 2, \ldots)\label{collision-eq1}
\end{align}
and let $u^0 = a \delta_{1, 0}$. Then, we obtain that a soliton $u^t = (-1)^{t} a\delta_{1, -t}$ for $t = 0, 1, 2, \ldots$, which moves to the left on the $x$-axis with speed $1$. It is easily seen from the definition of the map $S$ that the soliton moves to the right on the $x$-axis if $u^0 = a\delta_{2, 0}$. We call such soliton rotating soliton. On the other hand, a solution with $u^0 = a\delta_{1, 0}$ where $a$ is determined by solving 
\begin{align}
\dfrac{\pi}{4} + a^2 = 2n\pi \quad (n = 1, 2, \ldots) \label{collision-eq2}
\end{align}
becomes $u^t = a\delta_{1, -t}$, which keeps its shape and moves to the left on the $x$-axis with speed $1$. It moves to the right on the $x$-axis if $u^0 = \delta_{2, 0}$. We call this soliton traveling soliton. Furthermore, we regard the periodic solution corresponding to the equation \eqref{eq-Psoliton+} as a soliton since it is localized and its sup-norm does not decay for $t > 0$. We call such solution periodic soliton.

For the case with coin $C_-$, we have obtained the traveling soliton from \eqref{eq-Tsoliton-}. Other types of solitons are obtained by solving the following equations:
\begin{align}
&\dfrac{\pi}{4} -a^2 = -\dfrac{\pi}{2} - 2n\pi \qquad (n = 0, 1, 2, \ldots)&& \text{{\rm (periodic soliton)}}, \label{eq-Psoliton-} \\
&\dfrac{\pi}{4} -a^2 = -(2n+1)\pi \qquad (n = 0, 1, 2, \ldots)&& \text{{\rm (rotating soliton)}}. \label{eq-Rsoliton-}
\end{align}

Table 2 summarizes the values of $a$ obtained by \eqref{collision-eq1} and \eqref{collision-eq2} with $n = 1$ and by \eqref{eq-Psoliton-} and \eqref{eq-Rsoliton-} with $n = 0$ together with the cases \eqref{eq-Psoliton+} and \eqref{eq-Tsoliton-}. Figure \ref{soliton+RT} shows the behavior of rotating and traveling solitons for the case $C_+$. 
\begin{table}[h!]\label{table2}
\begin{center}
\begin{tabular}{|c|c|c|c|}
\hline
 & rotating soliton & traveling soliton & periodic soliton\\
\hline
$C_+$ & $a \approx 1.534990$ & $a \approx 2.344736$ & $a \approx 0.886227$\\
\hline
$C_-$ &  $a \approx 1.981664$ & $a \approx 0.886227$ & $a \approx 1.534990$\\
\hline
\end{tabular}
\caption{The values of $\|u^t\|_{l^\infty}$ of solitons, which are used in numerical studies in Section 4. }
\end{center}
\end{table}
Treating these kinds of solitons, we would like to consider the following four types of collision: 
\begin{enumerate}
\item[(I)] collision between the same solitons;

\item[(II)] collision between the rotating soliton and the traveling soliton;

\item[(III)] collision between the periodic soliton and the rotating soliton;

\item[(IV)] collision between the periodic soliton and the traveling soliton.
\end{enumerate}
It will turn out that Case I exhibits a simple dynamics which we can calculate explicitly, while the behavior of solutions after collision in Cases II--IV becomes complicated.   

\subsection{Collision: case I}
\begin{figure}[tbhp]
\begin{center}
\begin{tabular}{lll}
(a)\ Collision between rotating solitons \qquad \qquad &   (b)\ Collision between traveling solitons \vspace{0.8cm}\\
\includegraphics[width=6.0cm]{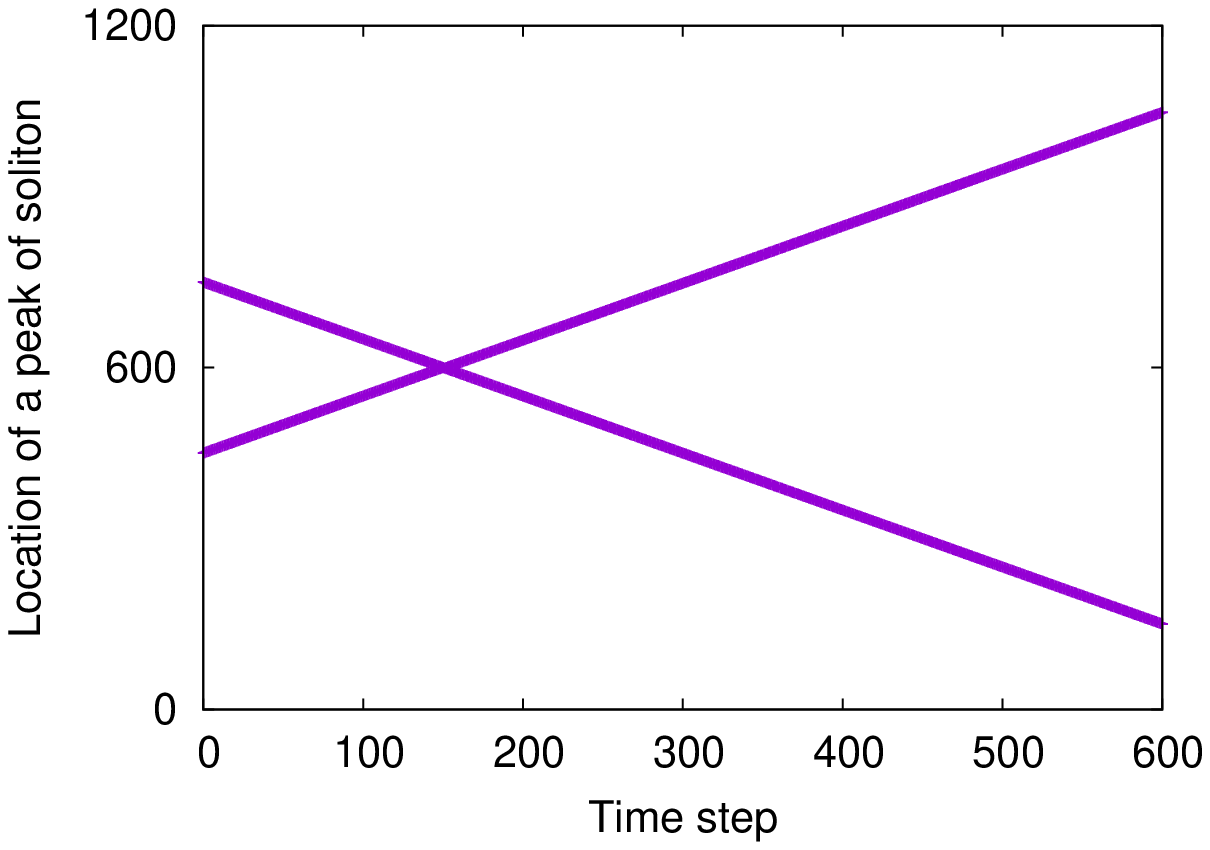}  
&
\includegraphics[width=6.0cm]{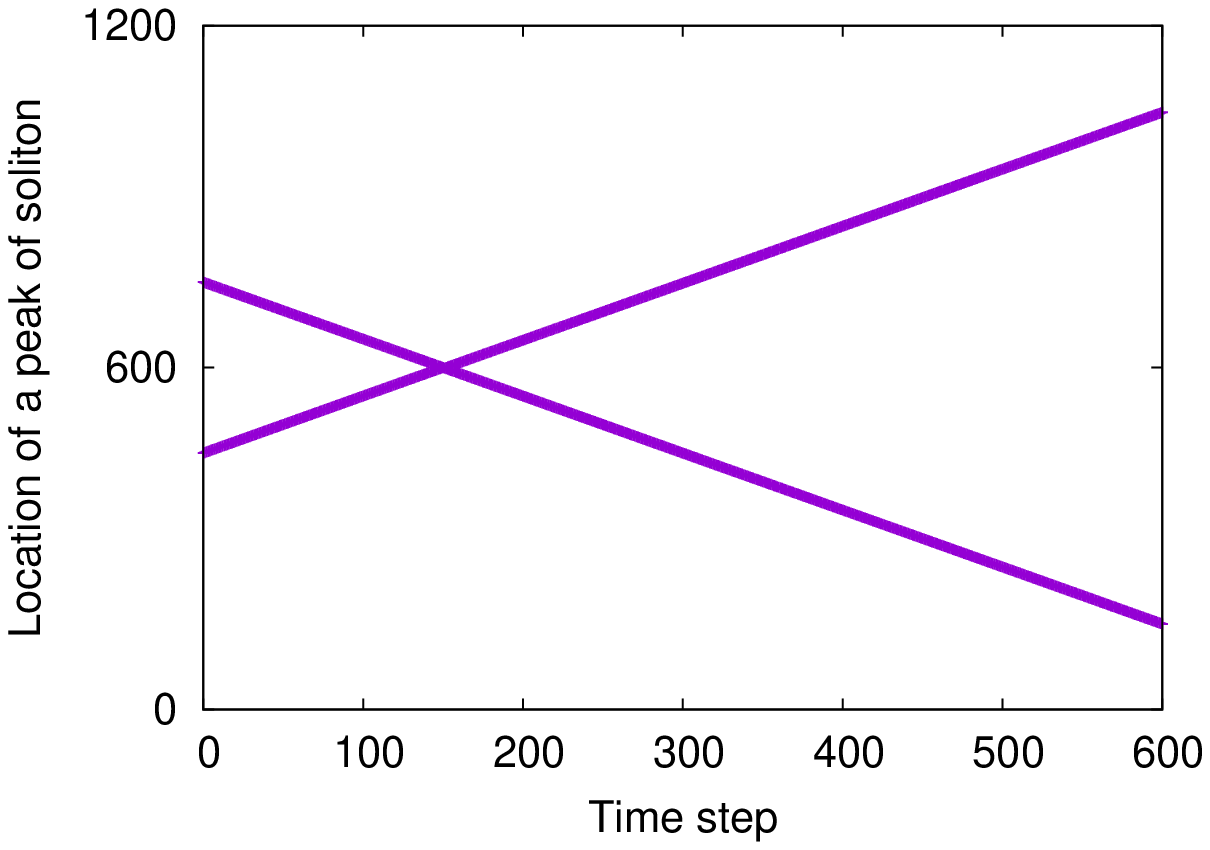} 
\end{tabular}
\end{center}
\caption{Location of the solitons $u^t$ with $C_+$. For each $t$, the points on $x$-axis with $u_1 \ge 0.3$ or $u_2 \ge 0.3$ are chosen, where $u^t = (u_1, u_2)^T$. Figure (a) shows a collision between rotating solitons. One moving to the right on the $x$-axis has its initial state $u^0 = 1.534990\delta_{2, 450}$ and another one moving to the left has its initial state $u^0 = 1.534990\delta_{1, 750}$. Figure (b) shows a collision between traveling solitons. One moving to the right on the $x$-axis has $u^0 = 2.344736\delta_{2, 450}$ and another one moving to the left has $u^0 = 2.344736\delta_{1, 750}$. }
\label{collision_2same_solitons}
\end{figure}

We consider two cases, one is the collision between rotating soliton and itself and another is the one between traveling soliton and itself. Since the case with $C_-$ exhibits exactly the same behavior, we discuss the case $C_+$ only.  

\begin{figure}[thbp]
\begin{tabular}{ccccccc}
 & &
\includegraphics[width=2.5cm]{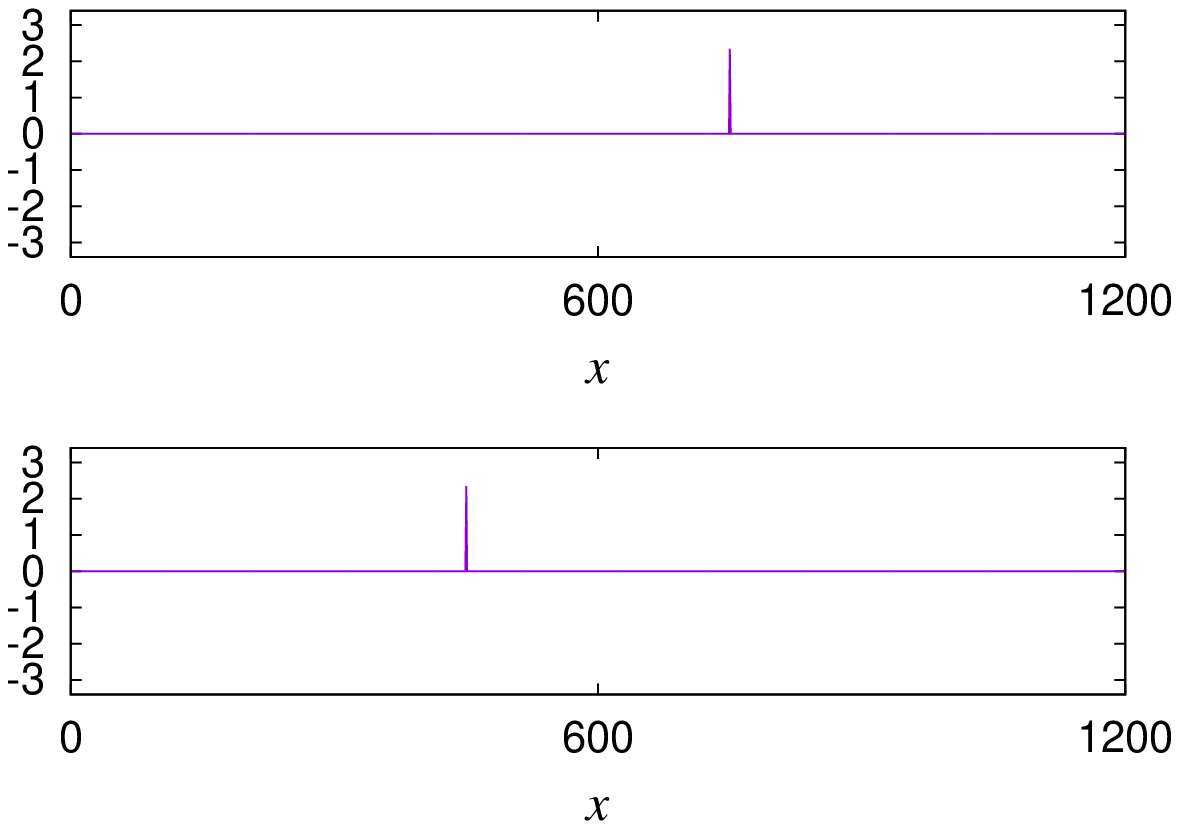}  &

\includegraphics[width=2.5cm]{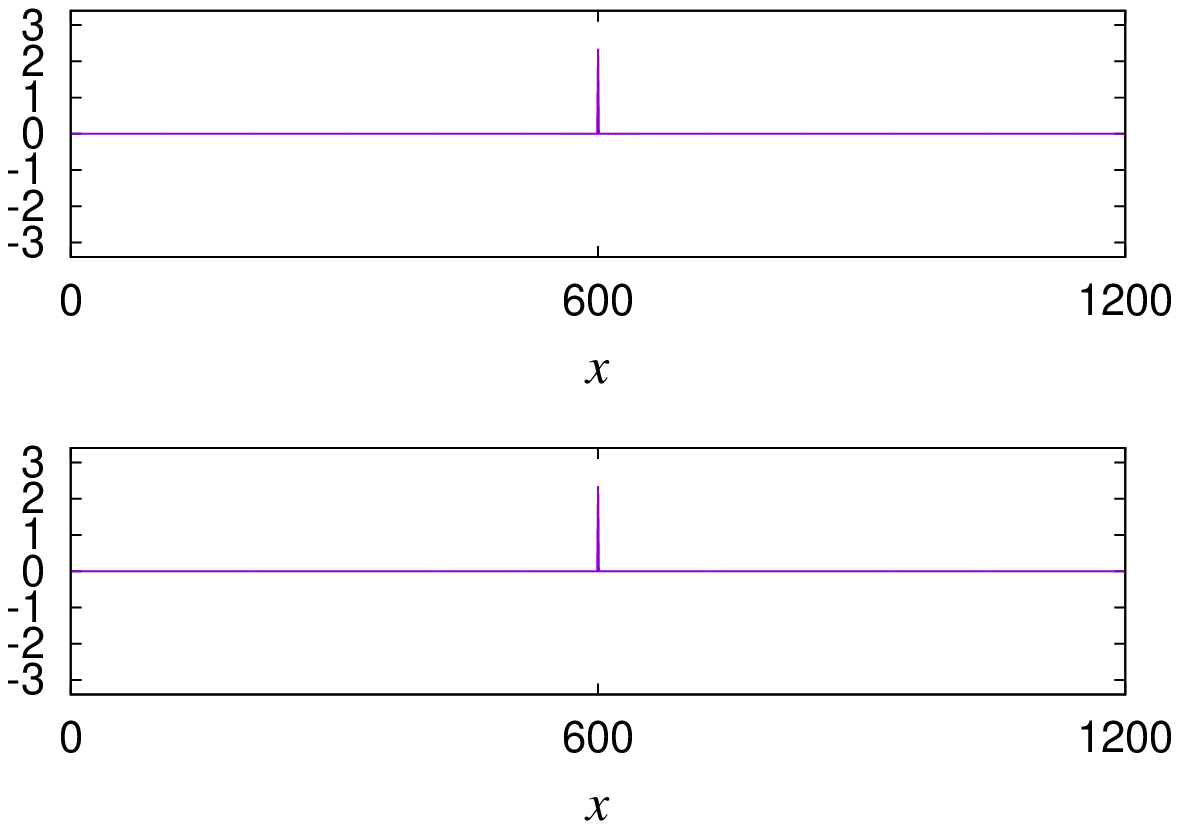}  &

\includegraphics[width=2.5cm]{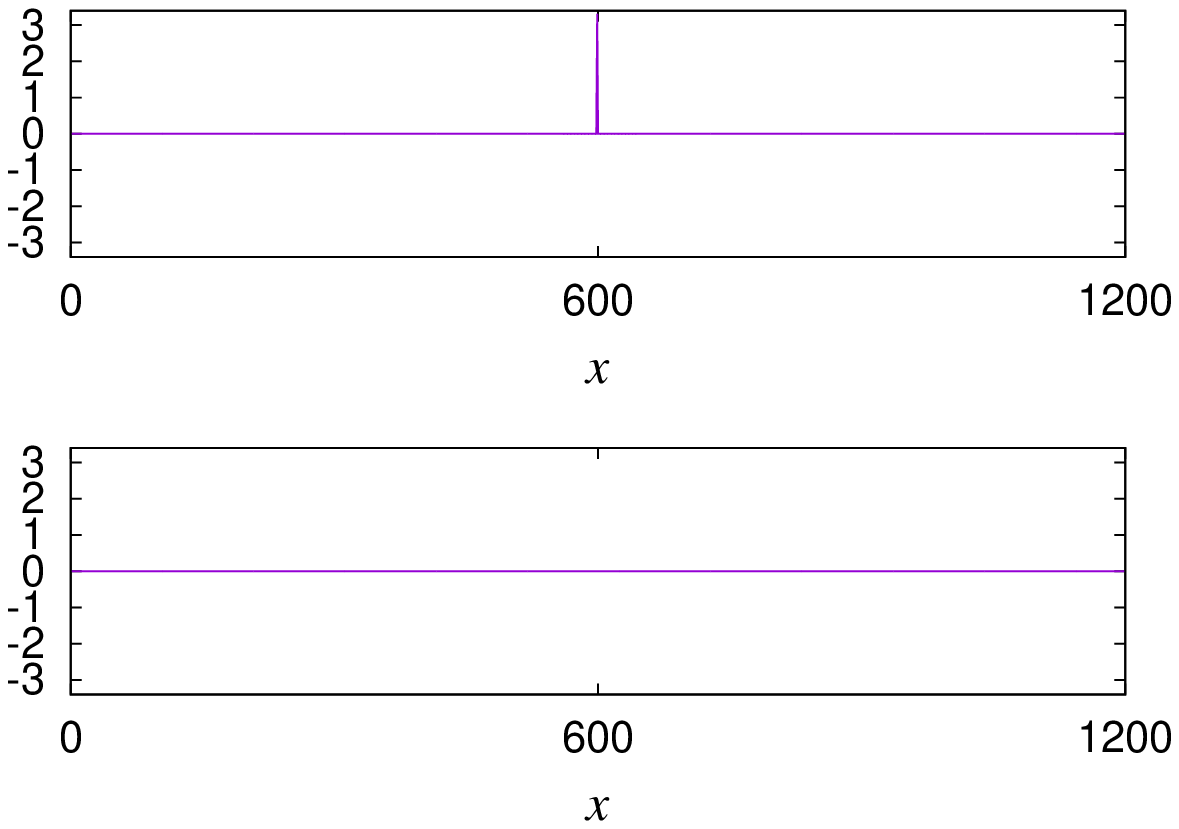}  &

\includegraphics[width=2.5cm]{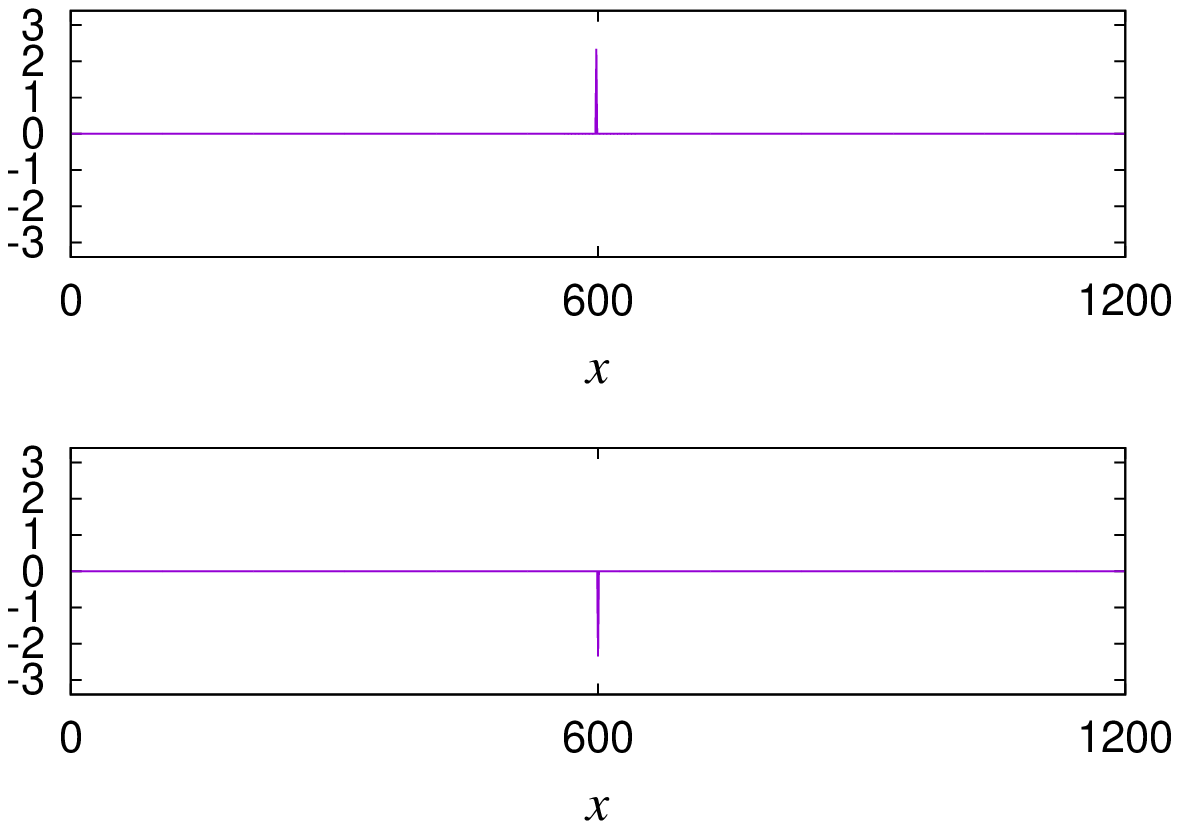}  &

\includegraphics[width=2.5cm]{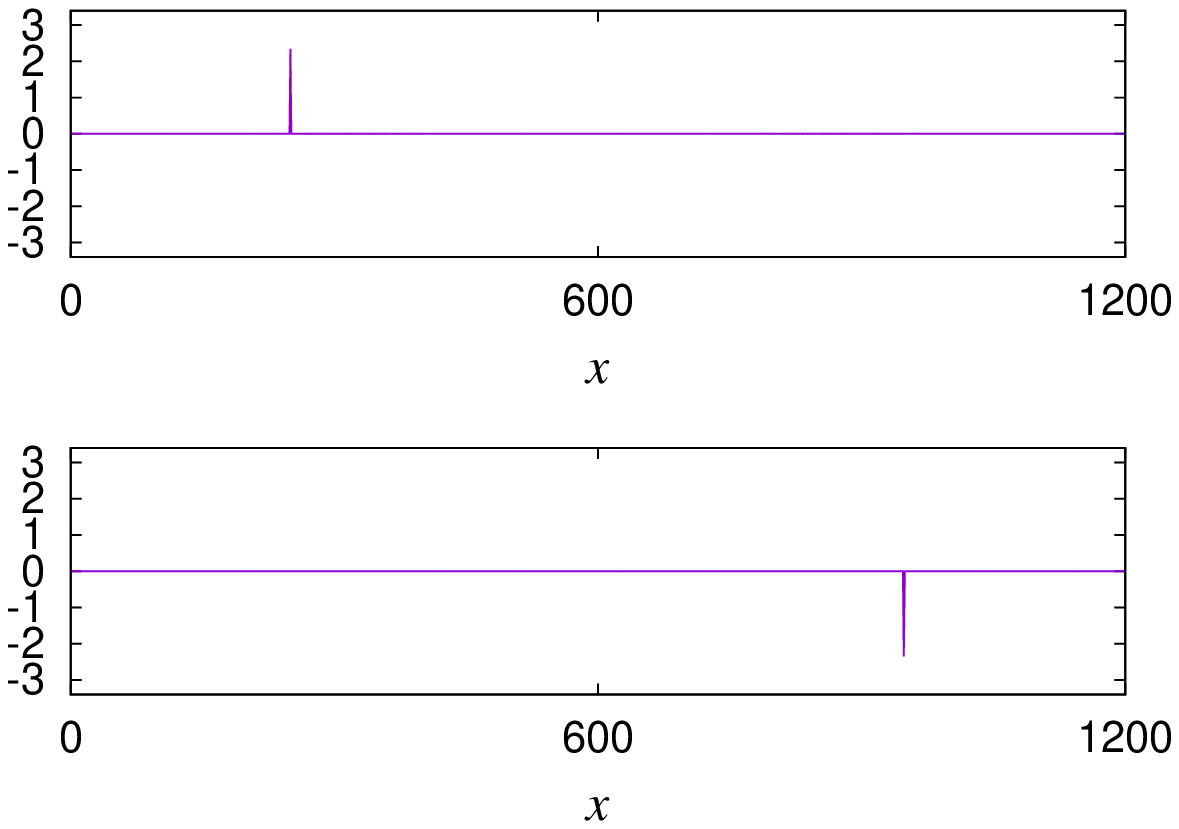} \\
 & & ${\footnotesize t = 0}$ & ${\footnotesize t = 150}$ & ${\footnotesize t = 151}$ & ${\footnotesize t = 152}$ & ${\footnotesize t = 500}$
\end{tabular}
\caption{Collision between traveling solitons. For each time step, the upper part shows the state of $u_1$ and the lower part is the state of $u_2$, where $u^t = (u_1, u_2)^T$. At $t = 0$, the initial state is given by $u^0 = 2.344736\delta_{1, 750}$ and $u^0 = 2.344736\delta_{2, 450}$. Since the traveling soliton has speed $1$, two solitons collide at $t = 150$. 
}
\label{collision_2same_solitons_2}
\end{figure}

Figure \ref{collision_2same_solitons} shows the location of solitons in the case $C_+$, where (a) is a collision between two rotating solitons and (b) is the one between two traveling solitons. It follows from these figures that location and speed of solitons do not change after collision. However, there is certainly interaction, which can be observed in Figure \ref{collision_2same_solitons_2}. Actually, we can explicitly calculate the behavior of $u^t$ at the collision time. 

Let $\alpha$ be the positive root of the equation \eqref{collision-eq2} with $n = 1$ and consider the collision between two traveling solitons. We demonstrate what happens in the behavior of $u^t$ with $u^0 = \alpha \delta _{1, 750} + \alpha \delta_{2, 450}$. Since the soliton with $u^0 = \alpha \delta _{1, 750}$ moves to the left and the one with $u^0 = \alpha \delta _{2, 450}$ moves to the right on the $x$-axis with their speeds $1$, they have no interaction for $0 \le t \le 149$, and we obtain 
\[
u^{150} = \alpha \delta_{1, 600} + \alpha \delta_{2, 600}.
\] 
Therefore, for $t = 151$, we see that
\begin{align*}
u^{151} &= U u^{150}\\
&=\left(
\begin{array}{cc}
T_- & 0 \\
0 & T_+
\end{array}
\right)
\left(
\begin{array}{cc}
\cos{\left(\dfrac{\pi}{4} + 2\alpha^2\right)} &  -\sin{\left(\dfrac{\pi}{4} + 2\alpha^2\right)} \vspace{0.3cm}\\
\sin{\left(\dfrac{\pi}{4} + 2\alpha^2\right)} & \cos{\left(\dfrac{\pi}{4} + 2\alpha^2\right)} 
\end{array}
\right)
\left(
\alpha \delta_{1, 600} + \alpha \delta_{2, 600}
\right).
\end{align*} 
Noting that $\pi/4 + \alpha^2 = 2\pi$ and $\alpha^2 = 2\pi -\pi/4$, we have
\begin{align*}
u^{151} 
&=
\left(
\begin{array}{cc}
T_- & 0 \\
0 & T_+
\end{array}
\right)
\sqrt{2}\alpha \delta_{1, 600} = \sqrt{2}\alpha \delta_{1, 599}.
\end{align*} 
We can continue the similar calculations for $t = 152$ to obtain that
\begin{align*}
u^{152} &= U u^{151}\\
&=\left(
\begin{array}{cc}
T_- & 0 \\
0 & T_+
\end{array}
\right)
\left(
\begin{array}{cc}
\cos{\left(\dfrac{\pi}{4} + 2\alpha^2\right)} &  -\sin{\left(\dfrac{\pi}{4} + 2\alpha^2\right)} \vspace{0.3cm}\\
\sin{\left(\dfrac{\pi}{4} + 2\alpha^2\right)} & \cos{\left(\dfrac{\pi}{4} + 2\alpha^2\right)} 
\end{array}
\right)
\sqrt{2}\alpha \delta_{1, 599}\\
&= \left(
\begin{array}{cc}
T_- & 0 \\
0 & T_+
\end{array}
\right)
\left(
\alpha \delta_{1, 599} -\alpha \delta_{2, 599}
\right) \\
&= \alpha\delta_{1, 598} -\alpha\delta_{2, 600}.
\end{align*} 
It is easily seen that there is no interaction for $t > 152$. Figure \ref{collision_2same_solitons_2} shows the behavior of $u^t$ which has been just proved.  
We note that the above behavior is very similar to the phenomena of collision for KdV solitons.
That is, if the two solitons did not interact with each other, then we should have
$\tilde u^{152}=\alpha \delta_{1,598}+\alpha \delta_{2,602}$.
So, comparing with $u^{152}$, we see that only the position and phase ($\alpha$ became $-\alpha$) have changed.

If we consider two traveling solitons but different values, namely, $u^t$ with 
\[
u^0 = \beta \delta_{1, 750} + \gamma \delta_{2, 450}, 
\]
where $\pi/4 + \beta^2 = 2n\pi$ and $\pi /4 + \gamma^2 = 2m\pi$\ $(n \neq m)$. Then, $u^t$ has the same behavior as the case $n = m= 1$ for $0 \le t \le 150$ and it becomes
\[
u^{150} = \beta \delta_{1, 600} + \gamma \delta_{2, 600}.
\]
However, for $t = 151$, we see that
\begin{align*}
u^{151} &= U u^{150}\\
&=\left(
\begin{array}{cc}
T_- & 0 \\
0 & T_+
\end{array}
\right)
\left(
\begin{array}{cc}
\cos{\left(\dfrac{\pi}{4} + \beta^2 + \gamma^2\right)} &  -\sin{\left(\dfrac{\pi}{4} + \beta^2 + \gamma^2 \right)} \vspace{0.3cm}\\
\sin{\left(\dfrac{\pi}{4} + \beta^2 + \gamma^2\right)} & \cos{\left(\dfrac{\pi}{4} + \beta^2 + \gamma^2\right)} 
\end{array}
\right)
\left(
\beta \delta_{1, 600} + \gamma \delta_{2, 600}
\right) \\
&= \left(
\begin{array}{cc}
T_- & 0 \\
0 & T_+
\end{array}
\right)
\left(
\dfrac{1}{\sqrt{2}}(\beta+\gamma)  \delta_{1, 600} + \dfrac{1}{\sqrt{2}}(\beta -\gamma) \delta_{2, 600}
\right) \\
&= \dfrac{\beta+\gamma}{\sqrt{2}}\delta_{1, 599} + \dfrac{\beta-\gamma}{\sqrt{2}}\delta_{2, 601}.
\end{align*} 
Since we cannot obtain the explicit values for $t > 151$, the behavior of $u^t$ becomes complicated. Consequently, using exactly the same solitons
is an important to obtain the simple dynamics of collision. 

Similar behavior is observed if we consider the collision between rotating solitons for $C_+$, and the same cases for $C_-$.

Comparing with the collision in Case I, the dynamics after collision in Cases II--IV is complicated. 

\subsection{Collision: case II}

\begin{figure}[tbhp]
\begin{tabular}{cc}
\begin{minipage}{0.45\hsize}
(a)\ $C_+$
\begin{center}
\includegraphics[width=6.0cm]{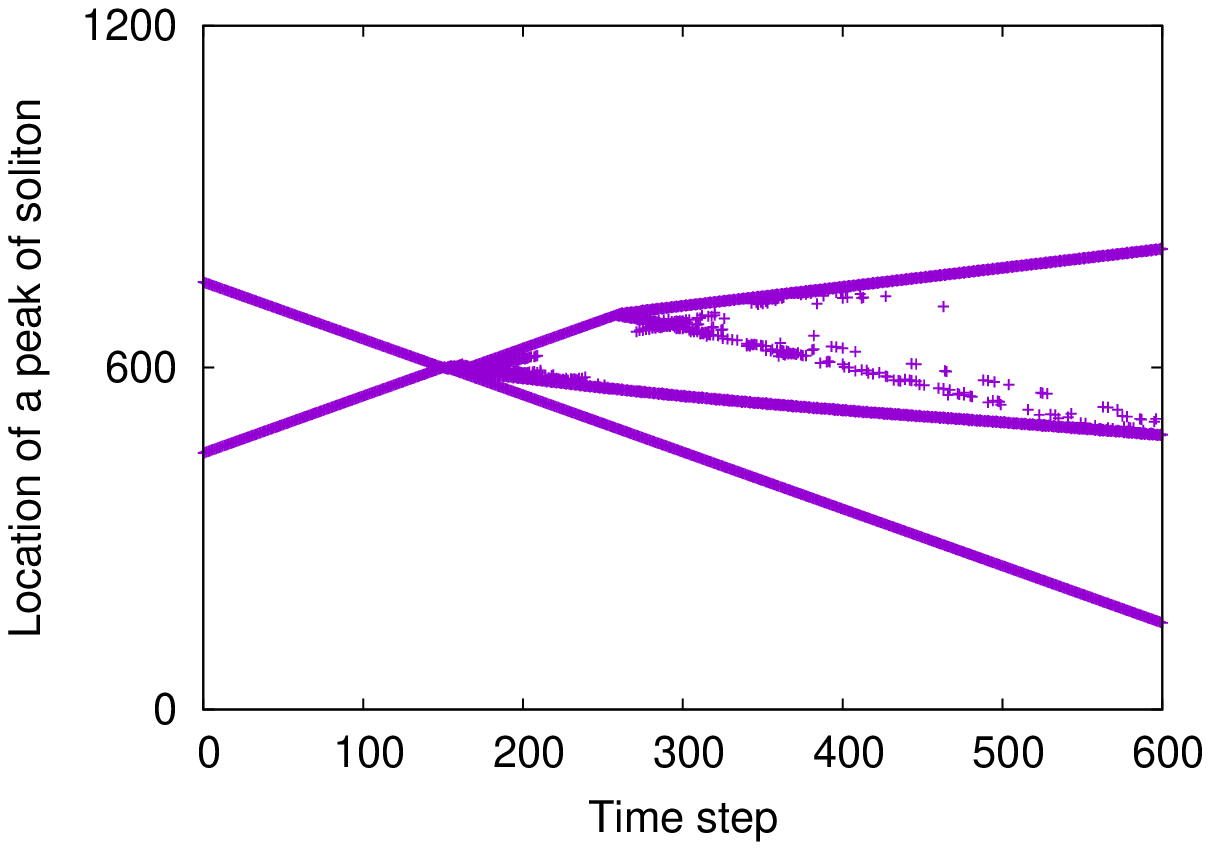}  
\end{center}
\end{minipage}
\hspace{6mm}
\begin{minipage}{0.45\hsize}
(b)\ $C_-$
\begin{center}
\includegraphics[width=6.0cm]{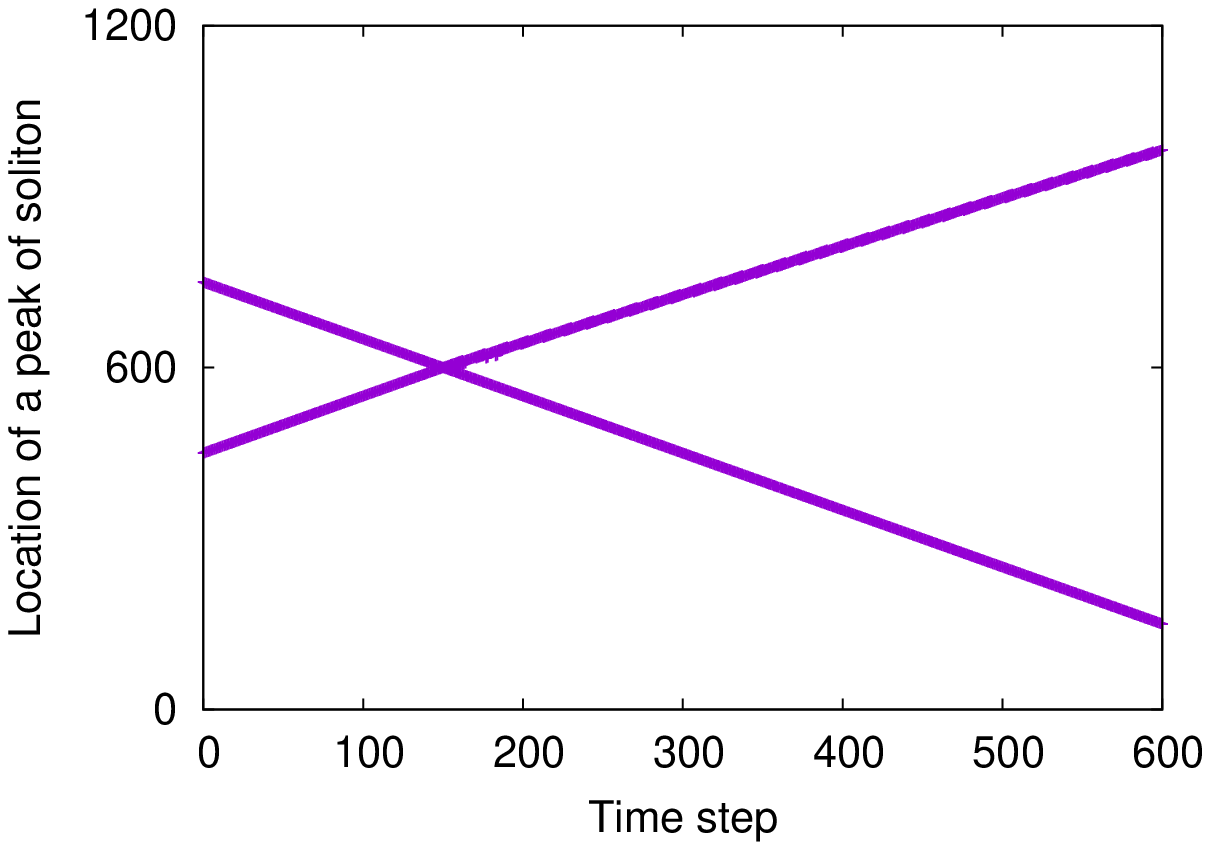} 
\end{center}
\end{minipage}
\end{tabular}
\caption{Collision between a rotating soliton and a traveling soliton. For each $t$, the points with $u_1 \ge 0.3$ or $u_2 \ge 0.3$ are chosen, where $u^t = (u_1, u_2)^T$. In both figures, the location of the peaks of a rotating soliton starting from $\delta_{1, 750}$ and of a traveling soliton starting from $\delta_{2, 450}$ are plotted.}
\label{collision_rotatingXtraveling}
\end{figure}

Figure \ref{collision_rotatingXtraveling} shows a collision between a rotating soliton and a traveling soliton for the case $C_+$ in (a) and for $C_-$ in (b).  Although Figure \ref{collision_rotatingXtraveling} (b) looks similar to Case I discussed in the previous subsection, it is not so simple and there should be interaction between two solitons. Indeed, we notice that the speed of soliton moving to the right on $x$-axis slightly changes after collision. 
We make an approximate calculation to understand the dynamics which observed in Figure \ref{collision_rotatingXtraveling} (b).

Let consider the behavior of $u^t$ for $C_-$ with its initial state $u^0 = \beta \delta_{1, 750} + \gamma \delta_{2, 450}$, where $\beta$ and $\gamma$ are positive roots of the equations \eqref{eq-Rsoliton-} and \eqref{eq-Tsoliton-}, respectively. 
Here, the part with $\beta \delta_{1, 750}$ corresponds to a rotating soliton and the one with $\gamma \delta_{2, 450}$ corresponds to a traveling soliton. Since the rotating and traveling solitons move with speed $1$, they collide at $t = 150$ to have
\[
u^{150} = \beta \delta_{1, 600} + \gamma \delta_{2, 600}.
\]
Then, for $t = 151$, we see that
\begin{align*}
u^{151} &= Uu^{150} \\
&=\left(
\begin{array}{cc}
T_- & 0 \\
0 & T_+
\end{array}
\right)
\left(
\begin{array}{cc}
\cos{\left(\dfrac{\pi}{4}-\beta^2 - \gamma^2\right)} &  -\sin{\left(\dfrac{\pi}{4}-\beta^2 - \gamma^2\right)} \vspace{0.3cm}\\
\sin{\left(\dfrac{\pi}{4}-\beta^2 - \gamma^2\right)}  & \cos{\left(\dfrac{\pi}{4}-\beta^2 - \gamma^2\right)} 
\end{array}
\right)
\left(
\beta \delta_{1, 600} + \gamma \delta_{2, 600}
\right).
\end{align*}
Noting that $\beta^2 = \pi /4 + \pi$ and $\gamma^2 = \pi /4$, we obtain that 
\begin{align*}
u^{151} &= \left(
\begin{array}{cc}
T_- & 0 \\
0 & T_+
\end{array}
\right)
\left(
-\dfrac{\beta + \gamma}{\sqrt{2}}  \delta_{1, 600} + \dfrac{\beta - \gamma}{\sqrt{2}} \delta_{2, 600}
\right) = -\dfrac{\beta + \gamma}{\sqrt{2}}  \delta_{1, 599} + \dfrac{\beta - \gamma}{\sqrt{2}}  \delta_{2, 601}.
\end{align*}
It follows that each peak at $x = 600$ moves one distance to the left or to the right, respectively. We should notice that each soliton loses its property since the value of the peak has been changed. 
For $t \ge 152$, the dynamics can be complicated because of the value of peaks. We obtain, for the peak on $x = 599$ at $t = 151$, that
\begin{align}
u^{152} &= U u^{151} \nonumber \\
&=\left(
\begin{array}{cc}
T_- & 0 \\
0 & T_+
\end{array}
\right)
\left(
-\dfrac{\beta + \gamma}{\sqrt{2}}\cos{\left(\dfrac{\pi}{4}-\dfrac{(\beta + \gamma)^2}{2}\right)} \delta_{1, 599} 
-\dfrac{\beta + \gamma}{\sqrt{2}}\sin{\left(\dfrac{\pi}{4}-\dfrac{(\beta + \gamma)^2}{2}\right)} \delta_{2, 599}
\right) \nonumber \\
&= -\dfrac{\beta + \gamma}{\sqrt{2}}\cos{\left(\dfrac{\pi}{4}-\dfrac{(\beta + \gamma)^2}{2}\right)}\delta_{1, 598}  -\dfrac{\beta + \gamma}{\sqrt{2}}\sin{\left(\dfrac{\pi}{4}-\dfrac{(\beta + \gamma)^2}{2}\right)} \delta_{2, 600}. \label{eq0306-1}
\end{align}
For the peak on $x = 601$ at $t = 151$, similar calculations lead to 
\begin{align}
u^{152} &= U u^{151}\nonumber \\ 
&=\left(
\begin{array}{cc}
T_- & 0 \\
0 & T_+
\end{array}
\right)
\left(
-\dfrac{\beta - \gamma}{\sqrt{2}}\sin{\left(\dfrac{\pi}{4}-\dfrac{(\beta - \gamma)^2}{2}\right)}\delta_{1, 601} + 
\dfrac{\beta - \gamma}{\sqrt{2}}\cos{\left(\dfrac{\pi}{4}-\dfrac{(\beta - \gamma)^2}{2}\right)} \delta_{2, 601}
\right) \nonumber \\
&= -\dfrac{\beta - \gamma}{\sqrt{2}}\sin{\left(\dfrac{\pi}{4}-\dfrac{(\beta - \gamma)^2}{2}\right)}\delta_{1, 600} + \dfrac{\beta - \gamma}{\sqrt{2}}\cos{\left(\dfrac{\pi}{4}-\dfrac{(\beta - \gamma)^2}{2}\right)} \delta_{2, 602}. \label{eq0306-2}
\end{align}
Since $\beta = \sqrt{5\pi}/2$ and $\gamma = \sqrt{\pi}/2$, we have $\beta \gamma = \sqrt{5}\pi/4$, where 
\[
\dfrac{\sqrt{5}}{4} \pi = \dfrac{1}{2}\pi + \theta \qquad \left(\theta = \dfrac{\sqrt{5}-2}{4}\pi \approx 0.059017 \pi \right).
\]
Noting that 
\[
\dfrac{\pi}{4}-\dfrac{(\beta + \gamma)^2}{2} = -(\pi + \theta), \qquad \dfrac{\pi}{4}-\dfrac{(\beta -\gamma)^2}{2}=\theta,
\]
we see that the main part of \eqref{eq0306-1} is the first term on the right-hand side and the one of \eqref{eq0306-2} is the second term on the right-hand side.   
Therefore, it follows from 
\[
\sin{\theta} \approx 0.184347, \qquad \cos{\theta} \approx 0.982861
\]
that $u^{152}$ looks like
\begin{align}
u^{152} \approx -\dfrac{\beta + \gamma}{\sqrt{2}}\delta_{1, 598} + \dfrac{\beta -\gamma}{\sqrt{2}}\delta_{2, 602}. \label{eq0306-3}
\end{align}
We note that 
\[
\dfrac{\beta + \gamma}{\sqrt{2}} \approx 2.279049, \qquad \dfrac{\beta -\gamma}{\sqrt{2}} \approx 0.774591.
\]
Hence, the left end part of $u^t$ on the $x$-axis behaves like a rotating soliton and converges to the one. On the other hand, traveling soliton was changed to an oscillating solution which was considered in Section 2.3. Then, the right end part of $u^t$ tends to $0$ and the part which is larger than $0.3$ emerges from interior part of $u^t$ (compare with Figure \ref{fig_oscillating} (c)). This shows the reason why the speed of the main part moving to the right has been changed in Figure \ref{collision_rotatingXtraveling} (b).

\subsection{Collision: cases III and IV}

\begin{figure}[tbhp]
\begin{tabular}{ll}
\begin{minipage}{0.45\hsize}
(a)\ $C_+$: periodic and a rotating solitons
\vspace{0.5cm}
\begin{center}
\includegraphics[width=6.0cm]{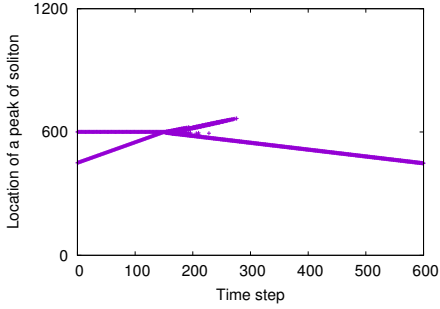}  
\end{center}
\end{minipage}
\hspace{6mm}
\begin{minipage}{0.45\hsize}
(b)\ $C_-$: periodic and a rotating solitons
\vspace{0.5cm}
\begin{center}
\includegraphics[width=6.0cm]{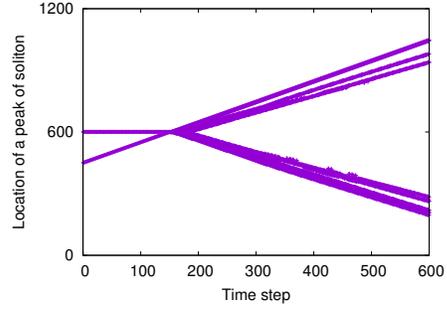} 
\end{center}
\end{minipage} \vspace{0.3cm}\\
\begin{minipage}{0.45\hsize}
(c)\ $C_+$: periodic and a traveling solitons
\vspace{0.5cm}
\begin{center}
\includegraphics[width=6.0cm]{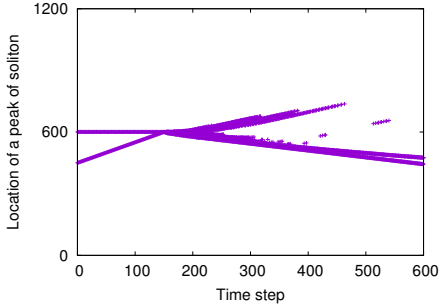}  
\end{center}
\end{minipage}
\hspace{6mm}
\begin{minipage}{0.45\hsize}
(d)\ $C_-$: periodic and a traveling solitons
\vspace{0.5cm}
\begin{center}
\includegraphics[width=6.0cm]{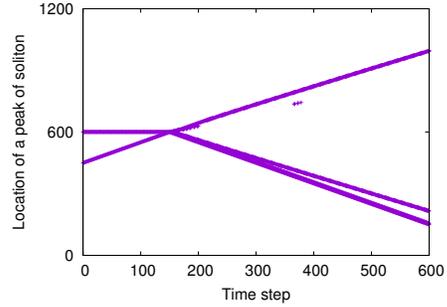} 
\end{center}
\end{minipage}
\end{tabular}
\caption{Collision between a periodic soliton and a rotating soliton ((a) for $C_+$ and (b) for $C_-$) and between a periodic soliton and a traveling soliton ((c) for $C_+$ and (d) for $C_-$). For each $t$, the points with $u_1 \ge 0.3$ or $u_2 \ge 0.3$ are chosen, where $u^t = (u_1, u_2)^T$. 
In each figure, the periodic soliton has an initial state $u^0 = a\delta_{1, 600}$, where $a = 0.886227$ for $C_+$ and $a = 1.534990$ for $C_-$. The rotating and traveling solitons start from $\delta_{2, 450}$.}
\label{collision_periodicX}
\end{figure}

Finally, we discuss the Cases III and IV. They exhibit similar dynamics after collision. Figure \ref{collision_periodicX} shows the complicated behavior of $u^t$. For the case $C_+$, a collision tends to lead a coalescence of soliton. On the other hand, several solitons appear after collision for the case $C_-$.

\section{Conclusion}\label{sec:disc}
In this paper, we have studied the dynamics of solutions of NLQWs in a strong nonlinear regime. If $u^t$ is localized, travels by a constant speed and $\|u^t\|_{l^\infty}$ does not decay for all $t > 0$, then we call such a solution a soliton solution. 
The existence of soliton solutions moving to the right or to the left with speed $1$ has been shown in Section 2, which consists of traveling and rotating solitons. In addition, there is a periodic soliton which evolves in a finite region with period $4$. 
Further, it is clear that one can construct a multi-soliton solution with all solitons moving to the right (or left) with constant speed $1$. 
We have systematically investigated a collision between two solitons choosing from a traveling soliton, a rotating soliton and a periodic soliton. Although it is expected that the dynamics becomes too complicated to analyze it, we could calculate the explicit process of collision between the same solitons (Subsection 4.1). Such behavior of solitons is very similar to the phenomena of collision for KdV solitons. 
On the other hand, we have also observed inelastic collision of solitons (Fig.\ 10 (a) and Fig.\ 11), which is a typical phenomenon of non-integrable PDEs such as a generalized KdV \cite{MM11AM}.
By this reason, we do not think the NLQWs which were considered in this paper are integrable.
However, this does not mean that all NLQWs are not integrable and we think it will be important to look for integrable NLQWs for deeper understanding of the dynamics of general NLQWs.


It would be interesting to study a large time behavior of oscillating solutions which has been obtained in Section 3.  We would like to know whether the oscillating solution decays for large $t$ or converges to some periodic solution. This is our future work.

\section*{Acknowledgments}  
M.M. was supported by the JSPS KAKENHI Grant Numbers JP15K17568, JP17H02851 and JP17H02853.
H.S. was supported by JSPS KAKENHI Grant Number JP17K05311.
E.S. acknowledges financial support from 
the Grant-in-Aid for Young Scientists (B) and of Scientific Research (B) Japan Society for the Promotion of Science (Grant No.~16K17637, No.~16K03939).
A. S. was supported by JSPS KAKENHI Grant Number JP26800054 and JP18K03327. 
K.S acknowledges JSPS the Grant-in-Aid for Scientific Research (C) 26400156 and 18K03354.

\medskip

Masaya Maeda, Hironobu Sasaki

Department of Mathematics and Informatics,
Faculty of Science,
Chiba University,
Chiba 263-8522, Japan

{\it E-mail Address}: {\tt maeda@math.s.chiba-u.ac.jp, sasaki@math.s.chiba-u.ac.jp}

\medskip

Etsuo Segawa

Graduate School of Information Sciences, 
Tohoku University,
Sendai 980-8579, Japan

{\it E-mail Address}: {\tt e-segawa@m.tohoku.ac.jp}

\medskip

Akito Suzuki

Division of Mathematics and Physics,
Faculty of Engineering,
Shinshu University,
Nagano 380-8553, Japan

{\it E-mail Address}: {\tt akito@shinshu-u.ac.jp}

\medskip

Kanako Suzuki

College of Science, Ibaraki University,
2-1-1 Bunkyo, Mito 310-8512, Japan

{\it E-mail Address}: {\tt kanako.suzuki.sci2@vc.ibaraki.ac.jp}

\end{document}